\documentclass[12pt]{article}

\usepackage[left=2cm,top=2.5cm,right=2cm,bottom=3cm,nohead]{geometry}
\usepackage{amsmath,amsfonts,amsthm,amssymb,times, bm}
\usepackage{epsfig}
\usepackage{color,verbatim,epic, eepic}
\input amssym.def
\input amssym.tex
\usepackage{yfonts}[1998/10/03]

\newfont{\go}{ygoth.tfm scaled 1200}  

\newtheorem{theorem}{Theorem}[section]

\newtheorem{prop}[theorem]{Proposition}

\newcommand{\abs}[1]{\left| #1 \right|}
\newcommand{\norm}[1]{\left\|#1\right\|}
\newcommand{\vect}[1]{\mathbf{#1}}

\newcommand{\Ht}{\mathbb{H}^3}

\newcommand{\dualA}{\widetilde{A}}

\newcommand{\Ga}[1]{\Gamma \left( #1 \right)}
\newcommand{\dif}[2]{\frac{d #1}{d #2}}
\newcommand{\pd}[2]{\frac{\partial#1}{\partial#2}}
\newcommand{\pdd}[2]{\frac{\partial^2 #1}{\partial #2 ^2}}

\newcommand{\db}[2]{ \left \{ {#1}, {#2} \right \} }

\newcommand{\com}[2]{ \left [ {#1}, {#2} \right ]}

\numberwithin{equation}{section}
\begin{document}

\begin{flushright}
\small
DAMTP-2006-72\\
27th Oct.\ 2006 \\
\normalsize
\end{flushright}

\begin{center}

\vspace{2cm}

{\LARGE {\bf Hidden symmetry of hyperbolic monopole motion \\}}

\vspace{1.5cm}

{\large G.\ W.\ Gibbons\footnote{G.W.Gibbons@damtp.cam.ac.uk} and C.\ M.\ Warnick\footnote{C.M.Warnick@damtp.cam.ac.uk}}

\vspace{.7cm}

\small
{Department of Applied Mathematics and
Theoretical Physics\\
Centre for Mathematical Sciences, University of Cambridge\\
Wilberforce Road, Cambridge, CB3 0WA, UK}
\vspace{2cm}

\thispagestyle{empty}
{\bf Abstract}

\end{center}

\begin{quotation}

\small

Hyperbolic monopole motion is studied for well separated monopoles. It
is shown that the motion of a hyperbolic monopole in the presence of
one or more fixed monopoles is equivalent to geodesic motion on a particular
submanifold of the full moduli space. The metric on
this submanifold is found to be a generalisation of the multi-centre
Taub-NUT metric introduced by LeBrun. The one centre case is analysed in
detail as a special case of a class of systems admitting a conserved Runge-Lenz
vector. The two centre problem is also considered. An integrable
classical string motion is exhibited.

\end{quotation}

\newpage
\section{Introduction}

One of the crowning achievements of classical mechanics was the
demonstration by Newton that Kepler's laws of planetary motion
followed from an inverse square law for gravity. The publication of Newton's {\it Philosophiae Naturalis
  Principia Mathematica} is considered by some to mark the beginning
of modern physics. Mathematically, as well as historically, this
problem is of particular interest. Bertrand's theorem \cite{Bertrand:1873} shows that the
only central forces that result in closed orbits for all bound
particles are the inverse-square law and Hooke's law. The reason for
these closed orbits in both cases is the existence of `hidden' or
`dynamical' symmetries. These are symmetries which only appear when
one considers the full phase space of the problem. In both cases the dynamical
symmetry group is larger than the na\"{i}ve $SO(3)$ arising from the
geometrical rotational symmetry of the central force problem. For the
inverse-square force, on whose generalisations we shall focus, the
dynamical symmetry group is $SO(4)$ in the case of bound states and
$SO(3,1)$ in the case of scattering states. The group is enlarged due
to the existence of an extra conserved quantity, known as the
Runge-Lenz vector although it pre-dates both Runge
and Lenz. Certainly Laplace and Hamilton both knew of this conserved
quantity but the identity of the first person to construct it is
unclear. It is claimed in \cite{Goldstein:1976} that the credit for
this belongs to Jakob Hermann and Johann I.\ Bernoulli in 1710, 89
years before Laplace and 135 years before Hamilton. Following common
usage however, we shall refer to the `Runge-Lenz' vector. Systems which admit such a
conserved vector are of great interest, and have been the subject of
much research over the years.

The `hidden' symmetries also appear in the quantum mechanical
problem. The system may be quantised by the usual process of replacing
Poisson brackets with commutators and in this way the energy levels
and degeneracies of the hydrogen atom may be found. The degeneracy is
greater than expected since the states of a given energy level fit
into an irreducible representation of $SO(4)$ rather than of
$SO(3)$. This explains why the energy depends only on the principle
quantum number and not on the total angular momentum as one might expect.

The conservation
of the Runge-Lenz vector also implies that the orbits of a particle moving
in a Newtonian potential are conic sections. Thus the mechanics of
Newton is elegantly related to the geometry of Euclid. In the
early 19th century Lobachevsky and
Bolyai independently discovered that other geometries are possible
which violate Euclid's parallel postulate: the spherical and
hyperbolic geometries. An interesting question is to what extent the
special properties of the Kepler problem in flat space carry over to
these new geometries. The generalization of the Kepler problem on spaces of
constant curvature was studied by Lipschitz \cite{Lipschitz} and Killing \cite{Killing} and later
rediscovered and extended to the quantum case by Schr\"{o}dinger
\cite{Schrodinger:1940xj} and Higgs \cite{Higgs:1978yy}\footnote{For a review of the history of this problem see \cite{Shchepetilov}}. It was found that a
Runge-Lenz vector may be constructed which is again conserved. Unlike
in Euclidean space, the dynamical symmetry algebra is not a Lie
algebra. The problem on hyperbolic
3-space $\Ht$ is closely related to that on $S^3$ and most results are valid for both
when expressed in terms of the curvature. The problem in hyperbolic
space is simpler in a sense, since global topological restrictions
prevent, for example, a single point charge from existing on the
sphere.

The proof of Bertrand assumes that the force on the particle depends
only on its distance from the origin. A particle moving in a magnetic
field will experience velocity dependent forces and so violates the
assumptions of the theorem. The simplest magnetic field that one may
consider is that due to a magnetic monopole at the origin. It was
discovered by Zwanziger \cite{Zwanziger:1969by} and independently by
McIntosh and Cisneros \cite{Mcintosh:1970gg} that a charged particle
moving in the field of a monopole at the origin has closed
orbits if the potential takes the form of a Coulomb potential together
with an inverse square term. This problem, known as the MIC-Kepler or
MICZ-Kepler problem can also be generalised to
the sphere and the hyperbolic space \cite{Nersessian:2000nv,Gritsev:1979, Kurochkin:2005}. There is once again a hidden symmetry
responsible for the closed orbits, which is a generalisation of the
Runge-Lenz vector. For a review of these systems, see \cite{Borisov:2005}.

A final system which exhibits a conserved Runge-Lenz vector is that of
two well separated BPS monopoles in $\mathbb{E}^3$. These are
solitonic solutions to $SU(2)$-Yang-Mills which behave like particles
at large separations. For slow motions, it is possible to describe the
time evolution of the field in terms of a geodesic motion on the space
of static configurations, the moduli space \cite{Manton:1981mp}. In
the case of two monopoles scattering, the full moduli space metric was
determined by Atiyah and Hitchin \cite{Atiyah:1985fd}. In the limit
when the monopoles are well separated, this metric approaches the
Taub-NUT metric with a negative mass parameter. The problem of
geodesic motion in Taub-NUT is in many ways similar to the motion of a
particle in a Newtonian potential. In particular, there is once again
a conserved vector of Runge-Lenz type which extends the symmetries
beyond the manifest geometric symmetries \cite{Gibbons:1986df, Feher:1986ib}. The orbits as a
consequence are conic sections and the energy levels of the
associated quantum system have a greater degeneracy than might be expected.

The question which we consider in this paper is to what extent the
results concerning well separated monopoles in flat space can be
extended to hyperbolic space. In section \ref{hypmon}
we show that for $SU(2)$ monopoles in $\Ht$ one can make progress by
assuming that some of the monopoles are fixed at given positions. The
motion is then described by a geodesic motion on a space whose metric
is a generalisation of the Taub-NUT metric where one considers a
circle bundle over $\Ht$ rather than $\mathbb{E}^3$. Such metrics were
first introduced by LeBrun \cite{LeBrun:1991}. In section \ref{lebrun}
we discuss some of the geometric properties of these metrics. In
section \ref{class} we consider the classical mechanics of a quite
general Lagrangian admitting a Runge-Lenz vector which includes as
special cases all of the systems mentioned above, in particular the
LeBrun metrics. Also included in
this section is a discussion of a particular classical string motion
in the LeBrun metrics which admits a Runge-Lenz vector. In section
\ref{quant} we consider the quantum mechanics and derive the energy
levels and scattering amplitude in the context of the general
Lagrangian. In section \ref{twocent} we consider the problem with two
fixed centres and show that this is integrable both in classical
and quantum mechanics, generalising a known result about the
integrability of the ionised hydrogen molecule. Finally we collect for reference in
the appendix some useful information about hyperbolic space.

\section{Hyperbolic Monopoles \label{hypmon}}

We are interested in $SU(2)$ monopoles on hyperbolic space. These are
defined as follows \cite{Austin:1990}. Let $\Ht$ be hyperbolic
$3$-space of constant curvature $-1$, with metric $h$. For a
principle $SU(2)$ bundle $P\to \Ht$, let $(A, \Phi)$ be a pair
consisting of a connection $A$ on $P$ and a Higgs field $\Phi \in
\Gamma(\Ht, g_P)$, where $g_P$ denotes the associated bundle of
Lie-algebras $P \times_{SU(2)}su(2)$. The pair $(A, \Phi)$ is a
magnetic monopole of mass $M \in \mathbb{R}_{>0}$ and charge $k \in
\mathbb{Z}_{\geq 0}$ if it satisfies the Bogomol'nyi equation
\begin{equation}
d_{A} \Phi = - \star_h F^{A} \label{bogom}
\end{equation}
and the Prasad-Sommerfeld boundary conditions
\begin{eqnarray}
M &=& \lim_{r\to\infty} \abs{\Phi(r)}, \nonumber \\
k &=& \lim_{r\to\infty} \frac{1}{4\pi} \int_{S_r^2} \mathrm{tr}\left(
\Phi F^{A}\right). \label{ps}
\end{eqnarray}
Here $S_r^2$ is a sphere of geodesic radius $r$ about a fixed point in $\Ht$
and $F^{A}$ is the curvature of the connection $A$. Hyperbolic
monopoles were introduced by Atiyah \cite{Atiyah:1987}, who
constructed them from $S^1$-invariant instantons on $S^4$. We define
the moduli space to be the space of solutions of (\ref{bogom},
\ref{ps}) modulo gauge transformations, i.e.\ the space of physically
distinct solutions. Atiyah also showed that the moduli space of hyperbolic monopoles of charge $k$ can be naturally identified with
the space of rational functions of the form
\begin{equation}
f(z) = \frac{a_1 z^{k-1}+a_2 z^{k-2}+\ldots + a_k}{z^k +
  b_1z^{k-1}+\ldots + b_k}
\end{equation}
where the numerator and the denominator have no common factor. Thus
the moduli space is a manifold of dimension $(4k-1)$, provided $k
\geq 1$. In fact this was only shown for the case when $2M$ is
an integer. Following \cite{Austin:1990} we shall assume that the moduli space is a manifold
of this dimension for all values of $M$. The moduli space is often
taken to be enlarged by a circle factor. This can be defined by fixing
a direction in $\Ht$, say $x_1$, choosing the gauge so that $A_1 = 0$
and then only considering gauge transformations which tend to the
identity as the hyperbolic distance along $x_1$ tends to
infinity. From now on we shall consider this enlarged moduli space,
which has dimension $4k$.

For BPS monopoles in flat space, it is possible to find a metric on
the moduli
space for well separated monopoles by treating them as point
particles carrying scalar, electric and magnetic charges
\cite{Manton:1985hs}. We denote by $\mathcal{M}_k$ the moduli space
of $k$-monopoles. The $k$-fold covering of $\mathcal{M}_k$,
$\widetilde{\mathcal{M}}_k$ splits as a metric product \cite{Atiyah:1988jp}.
\begin{equation}
\widetilde{\mathcal{M}}_k = \widetilde{\mathcal{M}}^0_k\times S^1
\times \mathbb{E}^3. \label{metsplit}
\end{equation}
The $\mathbb{E}^3$ component physically represents the
centre of motion of the system, which separates from the relative
motions governed by the metric on $\widetilde{\mathcal{M}}^0_k$. The
$S^1$ factor is an overall phase, which may be thought of as a total
electric charge which is conserved. The fact that the centre of mass
motion can be split off in this fashion is due to the Galilean
invariance of the theory, which appears as a low velocity limit of the
full Poincar\'{e} invariance of the $SU(2)$-Yang-Mills-Higgs
theory. In the case of two well separated monopoles
$\widetilde{\mathcal{M}}^0_2$ is Euclidean Taub-NUT with a negative
mass parameter. 

For
monopoles in $\Ht$, the dimension of the moduli space is known to be
the same as that for flat space. We call the moduli space of
$k$-monopoles on hyperbolic space $\mathcal{N}_k$ and its $k$-fold
covering  $\widetilde{\mathcal{N}}_k$.  Little is currently known
about the geometry of this moduli space, although it is known that the
$L^2$ metric one defines in the flat case is infinite. What this means
for monopole scattering is unclear. It is certainly plausible that the
slow motions are still described by a moduli space geodesic motion. We
give support to this idea below by showing that this is the case in the unphysical situation where $k-1$ monopoles are at
prescribed locations. We shall assume that the metric on the space of
such configurations arises as the restriction of a metric describing
the interaction of $k$ free monopoles, however our results do not rest
on this being the case.

We may separate off an $S^1$
factor, corresponding to the phase at infinity,  so we might expect a metric on  $\widetilde{\mathcal{N}}_k$ to
have a metric
product decomposition analogous to (\ref{metsplit}) of the form
\begin{equation}
\widetilde{\mathcal{N}}_k = \widetilde{\mathcal{N}}^0_k\times S^1
\times \mathbb{H}^3,
\end{equation}
however there is no analogue of the Galilei group for hyperbolic
space. Translational invariance guarantees a conserved total momentum,
but since there are no boost symmetries, this cannot be split off as a
separate centre of mass motion. In general if one considers two
particles in $\Ht$ interacting by a force which depends only on their
separation, one finds that the relative motion depends on the total
momentum of the system \cite{Shchepetilov:1997}, \cite{Maciejewski}. Thus, it is not
unreasonable to expect that the moduli space metric will {\it not} split up
in this fashion. This means that if we
wish to study the scattering of two hyperbolic monopoles, we will be
studying geodesic motion with respect to a 7-dimensional moduli space metric.

In order to make this problem more tractable, we shall consider a
simplified situation where one or more monopoles are fixed at given locations. This
is unphysical in the sense of $SU(2)$ monopoles since for well
separated monopoles,  the mass of each is not
a free parameter, it is determined by the other charges and so we
cannot consider an infinitely heavy monopole in order to fix it in one
place. Although this situation does not represent a genuine motion of
hyperbolic monopoles, it is nevertheless of interest as it gives an
insight into the interaction of two hyperbolic monopoles. We find that the
equations of motion may be put into the form of geodesic motion on a
generalisation of the multi-centre metrics, first written down by
LeBrun \cite{LeBrun:1991}.

A similar situation is considered by Nash \cite{Nash:2006qe}, who
has investigated singular monopoles on $\Ht$. Here one allows the solution
of the Bogomol'nyi equations to be singular at some points $p_i$, but
the nature of the singularity is fixed. Nash shows using twistor
theoretic methods that the natural metric for a charge $1$ singular
monopole is indeed a LeBrun metric. Unlike the metrics we find below
which generalize the singular negative mass Taub-NUT metrics, Nash
finds metrics which generalize the everywhere regular positive mass
Taub-NUT metrics.

The LeBrun metrics share a lot of the properties
of the multi-centre metrics, including integrable geodesic equations
for the one and two centre cases. This allows progress to be made
analytically and indeed makes the metrics worthy of study in their own
right. It is also worth noting that the multi-centre metrics may be
found as hyper-K\"{a}hler quotients of the asymptotic $k$ monopole
moduli space metrics \cite{Gibbons:1995yw}. We may conjecture that these LeBrun metrics may
arise from the full hyperbolic monopole moduli space metric by some
analogous construction.

\subsection{Motion of a test monopole \label{testpart}}

In order to simplify the problem of hyperbolic monopole scattering,
let us consider $n$ monopoles fixed in $\Ht$ at points $P_I$,
$I=1 \ldots n$. Following Manton \cite{Manton:1985hs}, we treat
these monopoles as point particles carrying an electric, magnetic and
scalar charge. This is possible since outside the core, the $SU(2)$
fields abelianise as in flat space. If the radius of the hyperbolic space is large compared
to the core of each monopole, the mass and scalar charge will be given
in terms of the other charges by the Euclidean expressions. We wish to consider a test particle moving in the field
of $n$ identical monopoles, so we take the electric and magnetic
charges to be $q$ and $g$\footnote{Unlike in the previous section, we
  do not take units such that the magnetic charge is an integer.} respectively
for all of the monopoles. The scalar
charge of a monopole with magnetic charge $g$ and electric charge $q$
is $(g^2+q^2)^{\frac{1}{2}}$. We wish to interpret the electric charge
as a momentum, and since in the moduli space approximation all momenta
must be small, we assume that $q$ is small.

As in the case of flat space, the electric and magnetic
fields $\bm{E}, \bm{B}$ are one-form fields on $\Ht$, defined in terms of
the field strength 2-form by:
\begin{equation}
F = dt \wedge \bm{E} + \star_h \bm{B}. \label{elecmag}
\end{equation}
The field-strength 2-form, away from $P_I$, satisfies the vacuum
Maxwell equations:
\begin{equation}
dF=0, \qquad d \star F = 0. \label{max}
\end{equation}
In the above equations, $\star$ is the Hodge duality operator for the
metric
\begin{equation}
ds^2 = -dt^2 + h, \label{sptmmet}
\end{equation}
where $h$ is the metric on $\Ht$, with Hodge duality operator
$\star_h$. Clearly, from (\ref{max}) there is a symmetry under $F \to
\star F$. This is the usual duality, well known in flat space, that for
a vacuum solution one can replace $(\bm{E}, \bm{B})$ with  $(\bm{B},
-\bm{E})$ and Maxwell's equations continue to hold. Because of this
duality it is possible to locally find two gauge potentials, $A$ and $\dualA$,
which are related to $F$ by
\begin{equation}
F = dA, \qquad F = \star d \dualA. \label{pots}
\end{equation}
We will require both these potentials, since a monopole moving in this
electromagnetic field will couple to both of them. The potential and dual
potential at a point $P$ due to a monopole at rest at a point $P_I$ is given
by:
\begin{eqnarray}
A &=& -\frac{q}{4\pi}V dt + \frac{g}{4\pi} \omega \nonumber \\
\dualA &=& - \frac{g}{4\pi}V dt - \frac{q}{4\pi} \omega, \label{potI}
\end{eqnarray}
where $V$ is a function on $\Ht$ and $\omega$ is a one-form on
$\Ht$, which are defined by:
\begin{equation}
V = \coth D(P, P_I) - 1, \qquad \star_h d \omega=dV, \label{vom}
\end{equation}
with $D(P, P_I)$ the hyperbolic distance between $P$ and $P_I$. Since
$V$ satisfies Laplace's equation on $\Ht$, $d \star_h d V=0$,
$\omega$ may be determined up to a choice of gauge (which we
will return to later). Using these equations, it is possible to check
that $dA = \star d \dualA$, which implies Maxwell's equations
(\ref{max}) for the 2-form $F$. Since Maxwell's equations are linear,
it is straightforward to write down the potential and dual potential
at a point $P$ due to all $n$ of the monopoles at points $P_I$. This
is still given by (\ref{potI}), but with $V$ and $\omega$ defined
by
\begin{eqnarray}
V &=& \sum_{I=1}^n \coth D(P, P_I) - n \nonumber \\
dV &=& \star_h d \omega. \label{vomall}
\end{eqnarray}
There is also a scalar field in this theory, the Higgs field, $\Phi$. The
value of $\Phi$ at the point $P$ due to the monopoles at $P_I$ is
given by
\begin{equation}
\Phi = \frac{(g^2+q^2)^{\frac{1}{2}}}{4\pi}V, \label{higgs}
\end{equation}
with $V$ as in equation (\ref{vomall}). We now consider another
monopole at the point $P$, with velocity vector $\bm{v} \in
T_P\Ht$. This monopole has magnetic charge $g$, electric charge
$q'$ and mass $m$. The Lagrangian for the motion of this monopole in a field with
potential $A$ and dual potential $\dualA$ and with Higgs field $\Phi$
is:
\begin{equation}
L = [-m+(q^2+q'^2)^{\frac{1}{2}}\Phi](-g(v,v))^{\frac{1}{2}} + q'
\langle A, v \rangle + g \langle \dualA, v \rangle, \label{lagpart}
\end{equation}
Where $g$ is the metric given by (\ref{sptmmet}), $v=(1,\bm{v})$ is
the 4-velocity of the particle and $\langle \cdot, \cdot \rangle$ is
the usual contraction between vectors and one-forms. This is the form
of the Lagrangian argued by Manton in \cite{Manton:1985hs},
generalised to hyperbolic space. 

As mentioned above, we are interested in slow
motions of the monopoles, such as might be described by a moduli space
metric approach. Thus, we will assume that $\bm{v}$, $q$ and $q'$ are
all small and expand to quadratic order in these quantities. We define
$\tilde{q}=q'-q$ and take the coordinates of $P$ to be $r^i$,
$i=1,2,3$, so that
$v^i=\dot{r}^i$. We also add a constant term $m-m\tilde{q}^2/2g^2$ to the
Lagrangian, which will not affect the equations of motion. The new
Lagrangian after the expansion is:
\begin{equation}
L = \left( \frac{1}{2}m-\frac{g^2}{8\pi}V\right
)\dot{r}^ih_{ij}\dot{r}^j + \frac{g\tilde{q}}{4\pi}
\omega_i\dot{r}^i - \frac{1}{g^2} \left( \frac{1}{2}m-\frac{g^2}{8\pi}V\right
) \tilde{q}^2. \label{modlag}
\end{equation}
We note that if we can interpret $\tilde{q}$ as a conserved momentum,
this Lagrangian will be quadratic in time derivatives and so may be
interpreted as a geodesic Lagrangian. In order
to do this, let us consider a circle bundle, $E$, over $\Ht \setminus \{
P_I\}$, which is the configuration space of our problem. We assume a
metric form:
\begin{equation}
ds^2 = A h + B (d \tau + W)^2
\end{equation}
with $h$ the metric on $\Ht$ and $W$ a connection on the
bundle. This gives a geodesic Lagrangian of the form:
\begin{equation}
L = \frac{1}{2}A \dot{r}^ih_{ij}\dot{r}^j + \frac{1}{2}B
\left(\dot{\tau}+\dot{r}^iW_i \right)^2. \label{efflag}
\end{equation}
Since $\tau$ is a cyclic coordinate, there is a conserved quantity
which we will identify as a multiple of $\tilde{q}$:
\begin{equation}
\tilde{q}=\kappa B(\dot{\tau}+\dot{r}^iW_i). \label{qtdef}
\end{equation}
As (\ref{efflag}) is a Lagrangian, we cannot simply use (\ref{qtdef})
to eliminate $\dot{\tau}$, but must instead use Routh's procedure (see
for example \cite{Goldstein})
which gives a new effective Lagrangian with the same equations of
motion as (\ref{efflag}). This new Lagrangian is given by:
\begin{equation}
L' = \frac{1}{2}A\dot{r}^ih_{ij}\dot{r}^j + \frac{\tilde{q}}{\kappa}
\dot{r}^i W_i - \frac{1}{2}\frac{\tilde{q}^2}{\kappa^2B}.
\end{equation}
Comparing this with (\ref{modlag}), we see that they are the same if
we identify
\begin{equation}
A = m-\frac{g^2}{4\pi}V, \qquad W_i=\frac{g \kappa}{4\pi}\omega_i,
\qquad \frac{1}{B}=\frac{\kappa^2}{g^2}A.
\end{equation}
We fix the value of $\kappa$ to remove the Misner string
singularities due to $\omega$, and this gives $\kappa = 4\pi /
g$. In order to clean up the form of the Lagrangian, we take units
where $m=4\pi$ and $g=4\pi p$ and remove an overall factor of $4\pi$. We
thus have that the slow motion of the monopole is equivalent to a
geodesic motion on a space with metric at $P$ given by:
\begin{equation}
ds^2 = V h + V^{-1} \left ( d \tau + \omega \right)^2 \label{multcent}
\end{equation}
Where $h$ is the metric on $\Ht$ and we redefine $V$ so that
\begin{equation}
V = 1 +p\left(n- \sum_{I=1}^n \coth D(P, P_I)\right). \label{Vfinal}
\end{equation}
Finally, we still have that $\omega$ is defined in terms of $V$ by the
relation
\begin{equation}
\star_h d \omega = d V. \label{homega}
\end{equation}
This only defines $\omega$ up to an exact form on $\Ht$, however
we can see from (\ref{multcent}) that a change of gauge for $\omega$ simply corresponds to a
coordinate transformation on $E$.

Thus we find that the motion of a monopole in $\Ht$ in the presence of
$n$ fixed monopoles may be described by geodesic motion on a
hyperbolic multi-centre metric given by (\ref{Vfinal}),
(\ref{homega}). Geometrically we may think of this as the asymptotic metric on a
submanifold within the full moduli space metric. This 
however will not be a geodesic submanifold since there is no mechanism to fix $n$ of
the monopoles while allowing one to move.

\section{The LeBrun metrics \label{lebrun}}

The metric (\ref{multcent}) was considered by LeBrun with $p=-1/2$ in 
\cite{LeBrun:1991}. For this choice of $p$, the metric is everywhere
regular, the apparent singularities at $P_I$ are nuts, i.e.\ smooth
fixed points of the $U(1)$ action generated by the Killing vector
$\pd{}{\tau}$. This can be seen by considering an expansion in a small
neighbourhood of one of the $P_I$ where the curvature of the
hyperbolic space can be neglected. The metric then looks like that of
Taub-NUT. In the case we consider above with positive $p$, the metric
will become singular near the fixed monopoles where $V$ becomes negative. Since we
wish to consider an approximation where all the monopoles are well
separated, we can ignore this singularity and we expect that it is
smoothed out in the full moduli space metric. This is precisely what
happens in the case of monopoles in flat space, where the full
Atiyah-Hitchin metric is regular, while the negative mass Taub-NUT is not.

LeBrun was interested in finding a metric on
$\mathbb{CP}_2 \# \cdots \# \mathbb{CP}_2$ which was K\"{a}hler and
scalar flat. He showed that in the conformal class of metrics of which
(\ref{multcent}) is a representative, there is a whole 2-sphere's
worth of different metrics which are scalar flat and K\"{a}hler, one
corresponding to each point at infinity of $\Ht$. More explicitly, he
showed that if $V$ and $\omega$ are as in (\ref{Vfinal}), (\ref{homega}) and
$q$ is any horospherical height function, then the metric
\begin{equation}
g = q^2(Vh+V^{-1}(d\tau+\omega)^2)
\end{equation}
is K\"{a}hler with scalar curvature zero. Here a horospherical height
function is a function on $\Ht$ whose restriction to some geodesic is
the exponential of the affine parameter and which is constant on the
forward directed horospheres orthogonal to this geodesic. The
horospheres are discussed in section (\ref{horo}). Since this
construction picks out a point on the boundary of $\Ht$ arbitrarily,
there is a two-sphere's worth of conformal factors which make the
metric (\ref{multcent}) K\"{a}hler.  These metrics have self-dual
Weyl tensor and so are sometimes referred to as
half-conformally-flat.

LeBrun in addition showed that the metric $g$ represents a
zero-scalar-curvature, axisymmetric, asymptotically flat K\"{a}hler
metric on $\mathbb{C}^2$ which has been blown up at $n$ points situated along a straight complex line, i.e.\
$n$ points have been replaced with $\mathbb{CP}_1$. This aspect of the metrics was
considered in \cite{Hartnoll:2004rv} where $\mathbb{C}^2$ with an
arbitrary number of points blown up was considered as a spacetime foam
for conformal supergravity.

Since the LeBrun metrics are all conformally scalar flat, an
interesting question is whether there is a conformal factor such that
the metric is locally Einstein. This was studied by Pedersen and Tod
\cite{Pedersen:1991} and they found that a metric of the form
(\ref{multcent}) is locally conformally Einstein if and only if $(V,
\omega)$ is a spherically symmetric monopole. In other words only the
one-centre metrics are conformally Einstein.

A K\"{a}hler form $K$ is both closed and covariantly constant
\begin{equation}
dK = 0, \qquad \nabla K = 0.
\end{equation}
Thus it necessarily satisfies Yano's equation:
\begin{equation}
dK = 3 \nabla K
\end{equation}
Returning to the conformal representative given by (\ref{multcent})
then, we find that this metric admits a 2-sphere's worth of {\it conformal}
Yano tensors which satisfy the conformal Yano equation:
\begin{equation}
K_{\lambda \kappa; \sigma}+K_{\sigma \kappa; \lambda} = \frac{2}{3}
\left(g_{\sigma \lambda}K^\nu{}_{\kappa;\nu}+g_{\kappa ( \lambda}K_{\sigma)}{}^\mu{}_{;\mu} \right).
\end{equation}
The existence of such conformal Yano tensors permits the construction
of symmetry operators for the massless Dirac equation on this space \cite{Benn:1996ia}.

We show below that in the case of the one- and two-centre metrics
there are additional higher rank symmetries which echo those of the
multi-Taub-NUT metrics. The one-centre metric has
Killing vectors which generate the $SU(2)\times U(1)$ isometry group
which can be made manifest by writing the metric in a Bianchi-IX
standard form. In addition, there are 3 rank two Killing tensors which
transform as a vector under the $SU(2)$ action and as a singlet under
the $U(1)$ action. These we interpret as a generalisation of the
Runge-Lenz vector. The two-centre metric has the reduced isometry
group of $U(1)\times U(1)$ and in addition admits a rank 2 Killing
tensor. This generalises the extra conserved quantity associated with
the motion of a particle in the gravitational field of two fixed
centres.

We see then that the LeBrun metrics have many attractive geometric properties
which are closely related to those of the multi-Taub-NUT metrics and
they are worthy of closer study.

\section{Classical mechanics of the one-centre problem \label{class}}

\subsection{Geodesics on Self-Dual Taub-NUT}
We begin by recalling some results about the geodesics of Self-Dual Taub-NUT
\cite{Gibbons:1986df, Gibbons:1986hz, Gibbons:1987sp}. Taub-NUT is a
four dimensional Hyper-K\"{a}hler manifold, with metric:
\begin{equation}
ds^2 = \left (1+\frac{p}{r}\right
)(dr^2+r^2(\sigma_1^2+\sigma_2^2))+p^2\left (1+\frac{p}{r}\right
)^{-1}\sigma_3^2, \label{taubnut}
\end{equation}
where $\sigma_i$ are the usual left invariant forms on $SU(2)$.
The geodesic Lagrangian for negative mass Taub-NUT (we take $p=-2$, since
changing $p$ only changes the overall scale of $L$) is
\begin{equation}
L = \frac{1}{2} \left [ \left( 1- \frac{2}{r} \right ) \vect{\dot{r}} \cdot
  \vect{\dot{r}} + 4 \left( 1- \frac{2}{r} \right )^{-1}(\dot{\psi} +
  \cos \theta \dot{\phi})^2 \right ] \label{taubnutlag}
\end{equation}
There is a conserved quantity $q$ associated to the ignorable coordinate
$\psi$ and also a conserved energy $E$. These are given by
\begin{equation}
q =  4 \left( 1- \frac{2}{r} \right )^{-1}(\dot{\psi} +
  \cos \theta \dot{\phi}), \qquad E =  \frac{1}{2} \left( 1- \frac{2}{r} \right ) \left (\vect{\dot{r}}^2 +
\left ( \frac{q}{2} \right )^2 \right). \label{taubnutcons}
\end{equation}
The conserved angular momenta are given by:
\begin{equation}
\vect{J} =   \left( 1- \frac{2}{r} \right )\vect{r} \times
\vect{\dot{r}} + q \vect{\hat{r}}, \label{taubnutang}
\end{equation}
where $\vect{\hat{r}}$ is a unit vector in the $\vect{r}$
direction. In addition to these conserved quantities, there is a extra
conserved vector, 
\begin{equation}
\vect{K} = \left ( 1-\frac{2}{r} \right) \vect{\dot{r}} \times
\vect{J} + \left (2 E - \frac{1}{2}q^2 \right)\vect{\hat{r}}. \label{taubnutrl}
\end{equation}
These conserved quantities allow all of the geodesics to be found.

As one would expect, the conserved quantities are related to geometric
symmetries of Taub-NUT. Taub-NUT has biaxial Bianchi-IX symmetry, so
the group of isometries is $SU(2) \times U(1)$. The $SU(2)$ factor
gives rise to the conserved vector $\vect{J}$ which transforms as a
triplet under the $SU(2)$ action. The $U(1)$ factor gives rise to the
conserved $SU(2)$ singlet $q$. The origins of the vector $\vect{K}$
are less obvious. Since $\vect{K}$ is quadratic in the momenta, it
arises from a triplet of rank 2 Killing tensors on Taub-NUT, each satisfying Killing's
equation:
\begin{equation}
K^i_{(\mu \nu ; \sigma)} = 0. \label{killing}
\end{equation}
These Killing tensors are best understood in terms of the K\"{a}hler
structures of Taub-NUT. Since Taub-NUT is Hyper-K\"{a}hler, there are
three independent complex structures $F^i_{\mu\nu}$. In addition,
Taub-NUT admits a further rank 2 Yano tensor, i.e. an anti-symmetric tensor
obeying Yano's equation:
\begin{equation}
Y_{\mu (\nu ; \sigma)}=0. \label{yano}
\end{equation}
The K\"{a}hler structures, being covariantly constant, also satisfy
Yano's equation trivially. It was shown in \cite{Gibbons:1987sp} that
the Killing tensors arise as the product of the complex structures
with the extra Yano tensor:
\begin{equation}
K^i_{\mu\nu} = Y_{\sigma(\mu}F^{i\sigma}{}_{\nu )}.
\end{equation}

\subsection{A class of systems admitting a Runge-Lenz vector}

We wish now to consider the generalisation of (\ref{taubnut}) which
describes the motion of a hyperbolic monopole about a fixed monopole,
as we derived in section \ref{testpart}. This corresponds to geodesic
motion on a manifold with metric (\ref{multcent}). We will find that
many of the integrability properties of geodesic motion in Taub-NUT
persist. Rather than start with the Lagrangian for geodesic motion, we
instead consider a more general Lagrangian and impose conservation of
a Runge-Lenz vector. We find that within the class of Lagrangians of
this form admitting a Runge-Lenz vector is the Lagrangian describing
geodesic motion on (\ref{multcent}). We shall consider the Lagrangian:
\begin{equation}
L = \frac{1}{2}\left [ V(\abs{\vect{r}}) \left (
  \frac{\dot{\vect{r}}^2}{1-\abs{\vect{r}}^2}+\frac{(\vect{r}\cdot\dot{\vect{r}})^2}{\left
  (1-\abs{\vect{r}}^2 \right)^2} \right) +
V(\abs{\vect{r}})^{-1}(\dot{\tau}+\bm{\omega} \cdot\dot{\vect{r}})^2 \right ] - W(\abs{\vect{r}}). \label{lagr}
\end{equation}
We note that the kinetic part is given by a metric of the form
(\ref{multcent}), with $h$ the metric on $\Ht$ in Beltrami coordinates
(see section \ref{beltrcoords}). We write this in spherical
coordinates, and assume a form for $\bm{\omega}$ such that the
Lagrangian (\ref{lagr}) admits a rotational  $SU(2)$ symmetry. Then:
\begin{equation}
L = \frac{1}{2} \left [ V(r) \left(
  \frac{\dot{r}^2}{(1-r^2)^2}+\frac{r^2}{1-r^2}(\dot{\theta}^2 + \sin^2\theta
  \dot{\phi}^2) \right) +V^{-1}(\dot{\tau}+p \cos \theta \dot{\phi})^2
  \right ] - W(r). \label{lagrsp}
\end{equation}
Since $\tau$ is cyclic, the conjugate momentum $p_\tau=e$ is
conserved. This is given by:
\begin{equation}
e=V^{-1}(\dot{\tau}+p \cos \theta \dot{\phi}). \label{taumom}
\end{equation}
There is also a conserved energy, $E$:
\begin{equation}
E = \frac{V(r)}{2} \left (
  \frac{\dot{r}^2}{1-r^2}+\frac{r^2}{(1-r^2)^2}(\dot{\theta}^2 + \sin^2\theta
  \dot{\phi}^2)  + e^2\right) + W(r). \label{energy}
\end{equation}
If we define a new variable $\psi$ by $\tau = p\psi$, then the
Lagrangian (\ref{lagrsp}) is manifestly $SU(2)$ invariant. Thus there
is a conserved angular momentum vector, which we can write as:
\begin{eqnarray}
\vect{J} &=& \frac{V(r)}{1-r^2}\vect{r} \times \vect{\dot{r}}+ e p
\vect{\hat{r}} \nonumber \\
&=& \vect{r} \times \vect{N} + e p \vect{\hat{r}}, \label{angmom}
\end{eqnarray}
where we have defined for convenience a new vector
\begin{equation}
\vect{N} = V(r) \frac{\dot{\vect{r}}}{1-r^2}. \label{Nmom}
\end{equation}

We would like to find the condition that the Lagrangian (\ref{lagr})
admits a Runge-Lenz vector in addition to the constants of the motion
above. By analogy with the Taub-NUT case, we posit a further conserved
quantity of the form:
\begin{equation}
\vect{K} = \vect{N}\times{\vect{J}}+g \hat{\vect{r}} \label{runge}.
\end{equation}
Using the Hamiltonian formalism it is a matter of straightforward, if
tedious, calculation to check that $\vect{K}$ is a constant of the
motion if and only if the functions $V(r)$, $W(r)$ and the constant
$g$ satisfy the equation
\begin{equation}
\frac{1}{2}e^2V(r)^2 + (W(r)-E)V(r) = \frac{g}{r} +
\frac{e^2p^2}{2r^2}+C, \label{rlcond}
\end{equation}
where $C$ is an arbitrary constant. Now $V(r)$ and $W(r)$ cannot
depend on $E$, as they are given functions in the Lagrangian
(\ref{lagr}). However $g$ and $C$ are free to depend on $E$ (see for
example (\ref{taubnutrl})). To find the most general functions
satisfying (\ref{rlcond}) for some $g(E)$, $C(E)$ we differentiate
twice with respect to $E$ and find:
\begin{eqnarray}
-V(r) &=& \frac{g'(E)}{r}+C'(E), \label{diffE} \nonumber \\
0 &=& \frac{g''(E)}{r} + C''(E). \label{diffEE}
\end{eqnarray}
From (\ref{diffEE}) we can write
\begin{equation}
g = -\alpha E + \beta, \qquad C = -kE + \gamma. \label{gdef}
\end{equation}
Substituting these back into (\ref{diffE}) and (\ref{rlcond}) we prove
the following proposition:

\begin{prop}
The most general functions $W(r)$, $V(r)$ such that the
Lagrangian (\ref{lagr}) admits a Runge-Lenz vector of the form
(\ref{runge}) are given by:
\begin{equation}
V(r) = k + \frac{\alpha}{r}, \qquad W(r) = \left (
\frac{\beta}{r}+\frac{e^2 p^2}{2 r^2} + \gamma \right )
\left(k+\frac{\alpha}{r}\right)^{-1}-\frac{1}{2}e^2
\left(k+\frac{\alpha}{r}\right). \label{VW}
\end{equation}
\end{prop}
From now on, we assume that $V(r)$ and $W(r)$ take this form.

\subsubsection{Some Special Cases}

We can find some special cases of the Lagrangian (\ref{lagr}) by
choosing particular values for the constants:

\begin{itemize}
\item If we take $k=1$, $\alpha=0$ and $e=0$, we have the Lagrangian for the
Kepler problem on hyperbolic space, $\Ht$.
\item If we take $k=1$, $\alpha=0$, but allow $e$ to be non-zero, we
  get the Lagrangian for the MICZ-Kepler problem in hyperbolic
  space \cite{Kurochkin:2005}. It will be useful for later to note that (formally at least)
  we can also obtain this system by setting $\alpha=p$ and letting
  $p\to 0$ while requiring that $e p = \mu$ remains constant.
\item If we take $\alpha=\mp p$, $\beta = -k\alpha e^2$, $\gamma =
  e^2k^2/2$ and $k=1\pm p$,  we have the Lagrangian for pure geodesic motion on the manifold
  with metric (\ref{multcent}). 
\end{itemize}

In specifying the metric on $\Ht$, we have everywhere assumed that the
space has radius $1$. Rather than carry an extra parameter through the
calculations, it is easier to replace the hyperbolic radius $R_0$ by
dimensional analysis in any equations. Doing this, we can consider
allowing $R_0$ to tend to infinity. The metric on $\Ht$ approaches
that on $\mathbb{E}^3$, so we might expect to recover known results
for flat space. In particular, with the special parameter choices
above, we may expect to recover results for the Kepler problem,
MICZ-Kepler problem and the problem of geodesic motion on Taub-NUT.

\subsubsection{Poisson Algebra of Conserved Quantities} 

We have 8 conserved quantities, $E$, $e$, $\vect{J}$, $\vect{K}$, 
which fall into two singlet and two triplet representations of 
$SU(2)$ respectively. We can calculate the non-vanishing Poisson 
brackets and after some algebra we find
\begin{eqnarray}  
\db{J_i}{J_j} &=& \epsilon_{ijk}J_k, \nonumber \\  
\db{J_i}{K_j} &=& \epsilon_{ijk}K_k,  \nonumber \\
\db{K_i}{K_j} &=& -2\epsilon_{ijk} J_k \left( Ek + J^2 - e^2 p^2  
-\gamma \right). \label{poisson}
\end{eqnarray} 
Thus the Poisson algebra of the conserved quantities is not a Lie
algebra, since the bracket of two components of $\vect{K}$ is a cubic
polynomial in
the conserved quantities. Algebras of this kind are known as finite
$W$-algebras. They have been extensively studied in
\cite{deBoer:1995nu}. This particular algebra is some deformation of
the $so(4)$ or $so(3,1)$ algebra, depending on the sign of
$Ek-e^2p^2 - \gamma$,  which leaves the $so(3)$ subalgebra undeformed. These
algebras are considered in \cite{Quesne:1995, Rocek:1991}.

\subsubsection{The shape of the orbits}

The existence of extra constants of the motion for the system defined
by (\ref{lagr}) and (\ref{VW}) means that it is possible to determine
the shapes of the orbits entirely from the conserved quantities. In
the Beltrami ball model (see \ref{beltrcoords}) we find that the orbits are given by conic
sections. 

From the definition of $\vect{J}$, (\ref{angmom}), we can immediately see
that
\begin{equation}
\vect{J} \cdot \vect{r} = e p r. \label{cone}
\end{equation} 
This is true for any functions $V(r)$ and $W(r)$ and it means that the
motion lies on a cone centred at the origin,  with axis parallel to  $\vect{J}$ and half angle $\theta$
satisfying $\cos \theta = e p /\abs{\vect{J}}$. In the case where
$ep=0$, we find that the motion is in a plane orthogonal to
$\vect{J}$. The $SU(2)$ symmetry therefore allows us to reduce the
three dimensional problem to motion on a two dimensional surface. In
general however, the orbits will not be closed and the motion may be
chaotic.

When we have a conserved Runge-Lenz vector however, any bound orbits
must be closed. To see this, we consider $\vect{K} \cdot \vect{r}$.
\begin{eqnarray}
\vect{K}\cdot\vect{r} &=& \vect{r}\cdot \vect{N}\times \vect{J} + gr,
\nonumber \\
&=& \vect{J} \cdot \vect{r} \times \vect{N} + g r, \nonumber \\
&=& \vect{J} \cdot \left(\vect{J} - e p \hat{\vect{r}} \right)+\frac{g}{ep}
\vect{J}\cdot\vect{r}. \label{kdotr}
\end{eqnarray}
Where we have used the definitions (\ref{angmom}), (\ref{runge}) and the
equation (\ref{cone}). Collecting these terms, we find that for $ep \neq 0$
\begin{equation}
\vect{r} \cdot \left(\vect{K}- \frac{g}{ep} \vect{J} \right) = J^2 -
e^2p^2. \label{plane}
\end{equation}
This is the equation of a plane orthogonal to $\vect{K}- \frac{g}{ep}
\vect{J}$ which in general does not pass through the origin. Thus the
orbits lie on the intersection of a cone and a plane and are hence
conic sections. 

If we take $ep=0$,
then $\vect{r}\cdot\vect{J}=0$ and  $\vect{K}\cdot\vect{J}=0$. The
motion is then in a plane perpendicular to $\vect{J}$ and we may take
the coordinates on this plane to be $r$ and $\theta$, the angle that
$\vect{r}$ makes with the vector $\vect{K}$. Taking $\vect{r}\cdot\vect{K}$
then yields the equation
\begin{equation}
r \left( 1- \frac{\abs{\vect{K}}}{g} \cos \theta \right) =
-\frac{J^2}{g} \label{consec}.
\end{equation}
This is the general form for a conic section in the plane. Thus, we
have shown that for the Lagrangian defined by (\ref{lagr}) and
(\ref{VW}) the general orbit is a conic section in Beltrami
coordinates. 

It should be noted that for $ep \neq 0$, the plane in which the motion
lies does not include the origin. Let us consider an incoming particle,
scattered by some infinitely heavy particle at the origin, such that
the motion is described by our Lagrangian. The two body system defines
a plane which includes the locations of both the particles and the velocity vector of
the moving particle. In the Kepler problem in flat space, the motion
of the particles is fixed in this plane, however here the plane of
motion receives a twist as the incoming particle is scattered. This phenomenon is noted in \cite{Gibbons:1986df} and is
typical of monopole scattering.

It is possible to consider all of the special cases mentioned above
and we find agreement with the results in the literature, as reviewed
in the introduction. We can also include the hyperbolic radius $R_0$,
and in the limit $R_0 \to \infty$ we recover the flat space results
for the Kepler and MICZ-Kepler problems, and also the results for
geodesic motion on self-dual Taub-NUT.

In conclusion, we have found that for the general Lagrangian admitting
a Rung-Lenz vector given above, the orbits are always conic sections
in an appropriate set of coordinates.

\subsubsection{The Hodograph}

In a paper of 1847 \cite{Hamilton:1847} Hamilton introduced the
hodograph, defined to be the curve traced out by the velocity vector,
thought of as a position vector from some fixed origin, as a particle
moves along its orbit. He showed that the Newtonian law of attraction
gave rise to a hodograph which is a circle, displaced from the origin,
and conversely that this was the only central force law to give a
circular hodograph. 

We shall define the hodograph in $\Ht$ to be the curve swept out by
the vector $\vect{N}$, thought of as a position vector originating at
$O$. By differentiating the condition (\ref{plane}) with respect to
time, and using the definition of $\vect{N}$, (\ref{Nmom}), it can be
seen that the hodograph lies in the plane
\begin{equation}
\vect{N} \cdot \left ( g \vect{J}-ep\vect{K} \right )=0. \label{Nplane}
\end{equation}
We may construct coordinates such that $\vect{N}$ lies in the
$x_1-x_2$ plane, then by squaring the definition of $\vect{K}$ it is
possible to show that the hodograph is an ellipse, with centre at
\begin{equation}
\frac{J^2K^2-e^2p^2g}{J^2\abs{\vect{J}\times\vect{K}}^2}\vect{J}\times\vect{K} \label{centre}
\end{equation}
and eccentricity given by
\begin{equation}
\epsilon^2 = ep\left(g - \frac{J^2(g^2-K^2)}{g(J^2-e^2p^2)}\right) \left(
\frac{e^2p^2}{g^2}\left(\frac{g^2-K^2}{J^2-e^2p^2}\right)^2 -2e^2p^2 \frac{g^2-K^2}{J^2-e^2p^2} +K^2\right)^{-\frac{1}{2}}.
\label{eccentricity}
\end{equation}
We see that in the case where $e$ or $p$ vanishes, the hodograph is
circular. In the case where $\alpha$ also vanishes, we find that the
Kepler problem does indeed give a circular hodograph. For $\alpha \neq
0$ we find a more general Lagrangian with a circular hodograph,
however this corresponds to modifying the kinetic term, so there is no
contradiction with Hamilton's result in flat space that the unique
central force law with a circular hodograph is the Newtonian
interaction.

\subsection{The Hamilton-Jacobi Method \label{actwav}}

An extremely powerful method for finding solutions to a problem in
Hamiltonian mechanics is the Hamilton-Jacobi method. For a
Hamiltonian $H(q^i, p_j)$, we seek solutions $S(q^i)$
to the Hamilton-Jacobi equation:
\begin{equation}
H(q^i, \pd{S}{q^j})=E. \label{hamjac}
\end{equation}
If a general solution to the Hamilton-Jacobi equation can be found
then it is possible to solve the system completely, i.e.\ give the
positions and momenta at a time $t$, as a function of $t$ and the
initial positions and momenta at some time $t_0$. On the other hand, a
particular solution of (\ref{hamjac}) defines a coherent family of
orbits whose momenta at each point are given by:
\begin{equation}
p_i = \frac{\partial S}{\partial q^i} \label{momact}
\end{equation}
Both these approaches to solutions of the Hamilton-Jacobi equation are
discussed by Synge \cite{Synge}. We will show that the Hamilton-Jacobi
equation separates, and then consider a particular solution
corresponding to Rutherford scattering in hyperbolic space.

The Hamiltonian for the
Lagrangian (\ref{lagr}), after replacing $p_\tau$ with $e$, may be written
\begin{equation}
H = \frac{1}{2} \left ( V^{-1} \left(p_i - e \omega_i
\right)h^{ij}\left(p_j - e \omega_j \right) + Ve^2 \right ) + W. \label{Hamiltonian}
\end{equation}

Since we have already solved the problem of motion governed by this
Lagrangian in Beltrami coordinates, we shall instead consider the
solution in pseudoparabolic coordinates (see section
\ref{ppara}). This will allow us to construct surfaces of constant
action for particles scattered by the origin. This was done in the
case of a Coulomb potential in flat space by Rowe \cite{Rowe:1985, Rowe:1987}.

We will require some results from section \ref{ppara}. Taking
cylindrical polars $\rho, z,\phi$ on the Beltrami ball, we can define
the pseudoparabolic coordinates $\mu, \nu$ to be given by:
\begin{equation}
z =\frac{\mu^2-\nu^2}{2+\mu^2-\nu^2}, \qquad \rho = \frac{\mu\nu}{2+\mu^2-\nu^2}. \label{munudef}
\end{equation}
Where 
\begin{equation}
\label{munuran}
0 \leq \mu < \infty, \qquad 0 \leq \nu < 1.
\end{equation}
In these coordinates the metric on $\Ht$ takes the form:
\begin{equation}
\label{munumetric}
ds^2 = \frac{\mu^2+\nu^2}{(1-\nu^2)(1+\mu^2)} \left [ \frac{d\mu^2}{1+\mu^2}+\frac{d\nu^2}{1-\nu^2}+\frac{d\phi^2}{\mu^{-2}+\nu^{-2}} \right ].
\end{equation}

We again assume that $\omega = p \cos \theta d \phi$. From the
equations (\ref{munudef}) and the standard relations between spherical
and cylindrical polar coordinates, it is a straightforward matter to
express $r$ and $\cos \theta$ in terms of $\mu$ and $\nu$. We then
have all of the necessary information to form the Hamilton-Jacobi
equation (\ref{hamjac}). We find that the equation separates if $W$
and $V$ are given by (\ref{VW}), with the additional condition that
$\alpha = p$, i.e.\ we require that equation (\ref{homega})
holds. Explicitly, we can write:
\begin{equation}
S = M(\mu)+N(\nu)+m \phi + e \tau, \label{addsep}
\end{equation}
then $M(\mu)$ and $N(\nu)$ satisfy the ordinary differential
equations:
\begin{eqnarray}
(1+\mu^2)\left(\frac{dM}{d\mu}\right)^2 + \frac{(m+ep)^2}{\mu^2} +
  \frac{2E(k-p)+2\beta-2\gamma-e^2p^2}{1+\mu^2} &=& K, \nonumber \\
(1-\nu^2)\left(\frac{dN}{d\nu}\right)^2 + \frac{(m-ep)^2}{\mu^2} -
  \frac{2E(k+p)+2\beta+2\gamma+e^2p^2}{1-\nu^2} &=& -K. \label{munuhamjac}
\end{eqnarray}
$K$ is an arbitrary constant which arises from the separation of
variables. It gives an extra constant of the motion related to the
Runge-Lenz vector. In the case of pure geodesic motion, we find that
$K$ is quadratic in the momenta, and so corresponds to a second rank Killing
tensor of the metric (\ref{multcent}). We have picked out a particular
direction in constructing our coordinates, so it is clear that there
is a rank 2 Killing tensor associated to each point on the sphere at
infinity.

The differential equations (\ref{munuhamjac}) can be solved
analytically, but for illustration we shall consider a simpler
case. As was previously noted, if we take the limit $p \to 0$, with
$e=0$ and $k=1$, we recover the Kepler (or Coulomb) problem on
hyperbolic space. We may set $\gamma=0$ without loss of generality,
since this simply corresponds to shifting the zero point of the
energy.  We do not seek to find the general solution of
(\ref{hamjac}). Following Rowe, we would like to find a solution which
describes a family of orbits, initially parallel to the positive
$z$-axis\footnote{We say that two geodesics are parallel if they meet
  on $\partial \Ht$. When we say a geodesic is parallel to the $z$-axis, we
  implicitly mean that it meets the positive $z$-axis on the boundary
  of $\Ht$. Two geodesics which do not meet, even on the boundary of $\Ht$
  are ultra-parallel.} and which are scattered by the Coulomb centre.

Since the particles move initially parallel to the $z$-axis, they have
no angular momentum about the $z$-axis, so we set $m=0$. We define
\begin{equation}
\kappa=\sqrt{2(E+\beta)} \qquad \mathrm{and} \qquad \lambda=\sqrt{2(E-\beta)}. \label{kaplam}
\end{equation}
The equations for $M$ and $N$ then simplify considerably, and we find
\begin{equation}
(1+\mu^2)\left(\frac{dM}{d\mu}\right)^2 = K - \frac{\kappa^2}{1+\mu^2},
  \qquad (1-\nu^2)\left(\frac{dN}{d\nu}\right)^2 = \frac{\lambda^2}{1-\nu^2}-K.
\end{equation}
We shall consider the case of an attractive Coulomb centre, i.e.\
$\beta \leq 0$.\footnote{The repulsive case follows in a similar
  fashion to the attractive case} We take $K=\lambda^2$ in order to
simplify the second equation. This is necessary in order to have the
correct $\nu$ dependence as $\mu \to \infty$, i.e.\ far from the origin
along the positive $z$-axis. Integrating then gives:
\begin{eqnarray}
N &=& \frac{ \lambda}{2}\log (1-\nu^2) \nonumber\\
M &=& \frac{\mp \kappa}{2} \log
\left(\frac{\sqrt{\lambda^2(1+\mu^2)-\kappa^2}+\kappa\mu}{\sqrt{
    \lambda^2(1+\mu^2)- \kappa^2}-\kappa\mu}\right ) + \lambda \log \left(
\pm \lambda \mu + \sqrt{\lambda^2(1+\mu^2)-\kappa^2} \right)
\label{actionsol}
\end{eqnarray} 
There are in principle two choices of sign to be made, one for
$\kappa$ and one for $\lambda$, however it can be seen that $\kappa
\to -\kappa$ does not change the equations (\ref{actionsol}). If we
consider the solution with the upper signs, as $\beta \to 0$ we find
that 
\begin{equation}
S(\mu, \nu) \sim \frac{\sqrt{2E}}{2}\log(1+\mu^2)(1-\nu^2) + f(\kappa,
\lambda) \label{freeact}
\end{equation}
The function $f$ is singular in the limit $\kappa \to \lambda$,
however $S$ is only defined up to an additive constant, so it can be
removed. The level sets of the function $(1+\mu^2)(1-\nu^2)$ are known
as horospheres and play a special r\^{o}le in scattering in
hyperbolic space. They are the hyperbolic equivalent of the plane
wave-fronts in flat space, see section \ref{horo}.

\begin{figure}[!t]
\begin{minipage}[t]{0.46\linewidth} 
\centering \framebox {\includegraphics[height=3in,
width=3in]{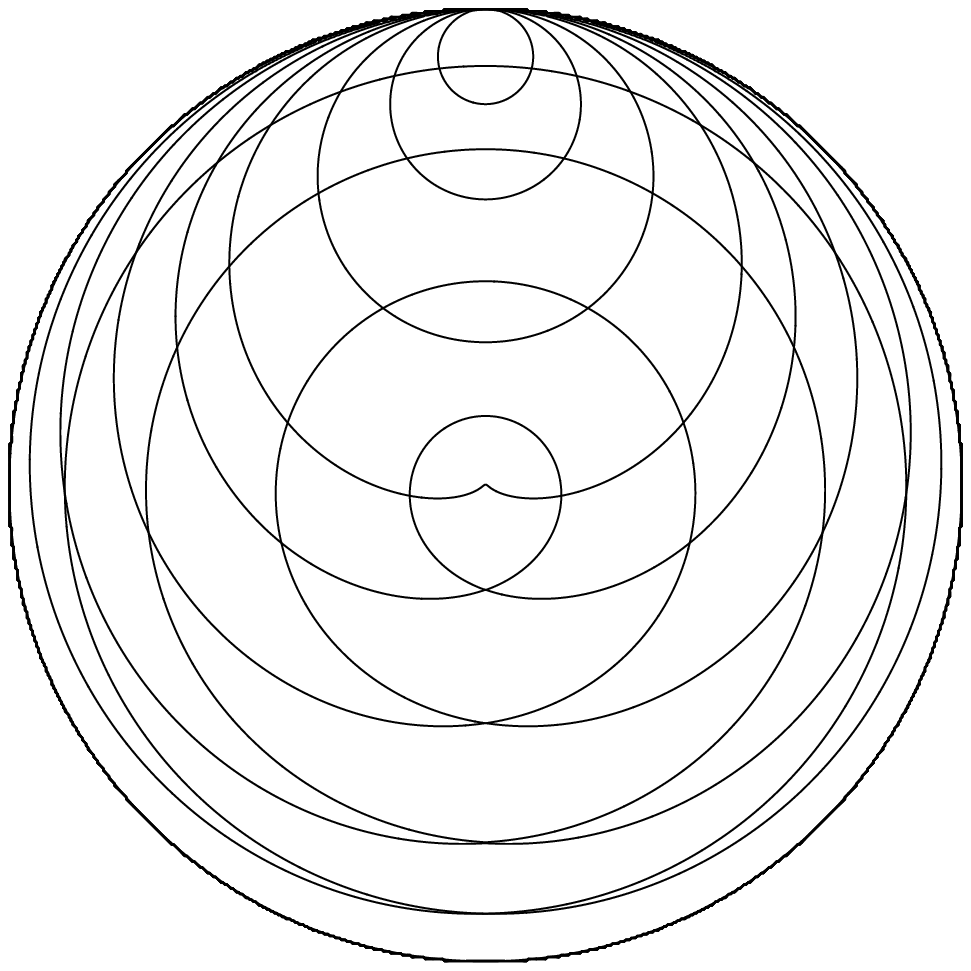}}
\caption{A plot showing the surfaces of constant action with $E=10$,
  $\beta=-1$ for the whole of $\Ht$ \label{confcirc}}
\end{minipage}
\hfill 
\begin{minipage}[t]{0.46\linewidth}
\centering \framebox {\includegraphics[height=3in,
width=3in]{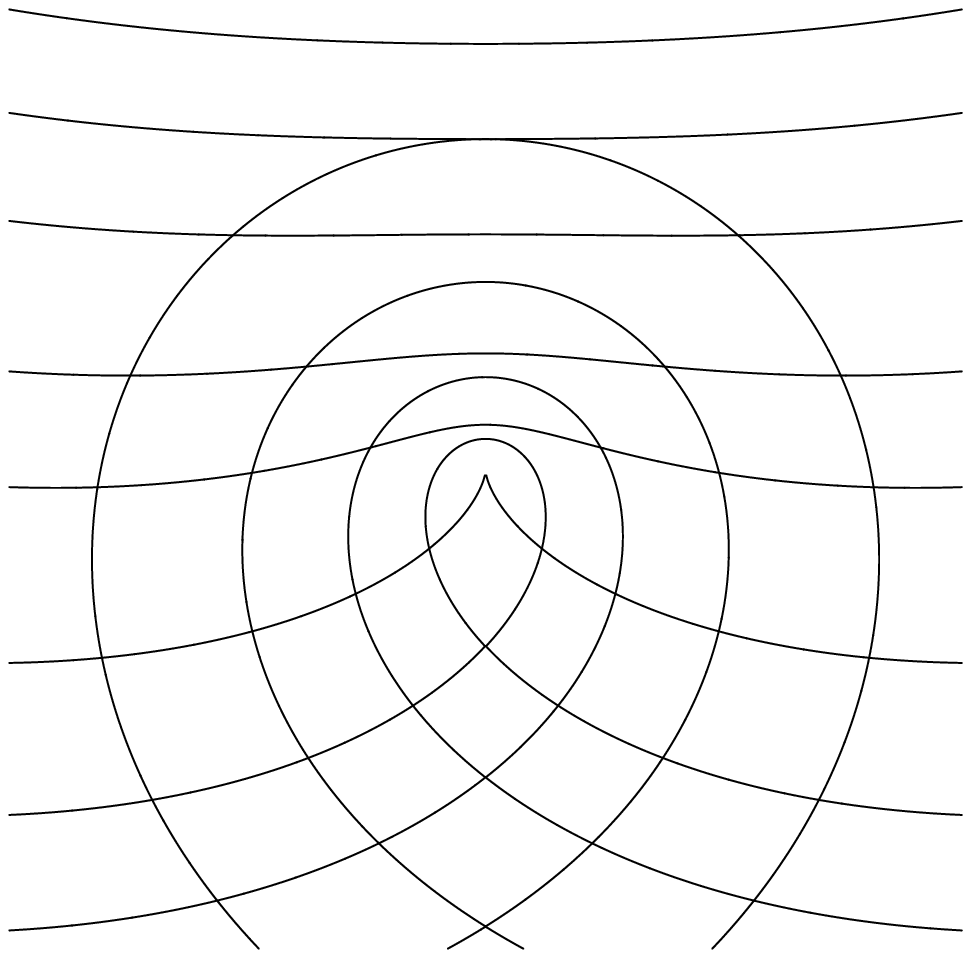}}
\caption{A plot showing the surfaces of constant action with $E=10$,
  $\beta=-1$ in the region around the origin \label{confloc}}
\end{minipage}
\end{figure}

The solution (\ref{freeact}) represents a family of free particles
which move parallel to the $z$-axis. If we allow $\beta$ to be
non-zero, (\ref{actionsol}) represents a family of particles
initially moving parallel to the $z$-axis but which are scattered by a
Coulomb centre at the origin. We need to consider the choice of sign
however. In fact, we need both branches of (\ref{actionsol}) to
describe the scattering process. Since $\Ht \setminus \{0\}$ is not
topologically trivial, we may allow multiple valued solutions to
(\ref{hamjac}). Alternatively, since both branches are equal on
$\mu=0$, we can follow Rowe and define $S$ as a
single valued function on two copies of hyperbolic space,
communicating through the negative $z$-axis. A particle which starts
initially parallel to the positive $z$-axis and is bent through the
negative $z$-axis will pass from one copy of $\Ht$ to the other.

 Taking $\mu \to \infty$, we see that the
Coulomb action approaches the free action, but with some distortion of
the plane wave. This is typical of the Coulomb potential, even in flat
space.

In order to see physically what these solutions mean, it is convenient
to plot the surfaces of constant action. For the Hamiltonian above,
with the parameters as defined, Hamilton's equations give
\begin{equation}
\dot{r}^j = h^{ij}p_j \label{hameq}
\end{equation}
thus the orbits are everywhere normal to the surfaces of constant
action. If we plot these surfaces in the Poincar\'{e} model of $\Ht$
(see \ref{confflat}), then the
orbits will cross the surfaces of constant $S$ at right angles in the
Euclidean sense. Figures \ref{confcirc} and \ref{confloc} show plots
in Poincar\'{e} coordinates
of the curves $S=\mathrm{const.}$ in a plane containing the $z$-axis. Both branches
of $S$ have been plotted on the same graph, and we see that they match
along the branch cut. We also see that one branch represents incoming
waves and the other gives scattered wavefronts.

\subsection{A classical string motion}

In \cite{Gibbons:1988xf} classical Nambu strings were studied on a multi-monopole
background of the form
\begin{equation}
g_{AB}dx^A dx^B = -dt^2+V^{-1}(dx^4+\omega_i dx^i)^2 + V h_{i
  j}dx^idx^j \label{5met}
\end{equation}
with $A,B=0,\ldots,4,$ $i,j=1, 2, 3$. Here $h_{ij}=\delta_{ij}$ is the metric on
$\mathbb{E}^3$, $x^4$ is a compact direction such that $0\leq x^4 \leq
2\pi R_K$, and $V$ and $\omega$ take the usual form:
\begin{equation}
V = 1 + \frac{R_K}{2}\sum_{i=1}^k \frac{1}{\abs{\vect{x}-\vect{x}_i}},
\qquad d\omega= \star_h dV.
\end{equation}
It was found that a classical string winding the $x^4$ direction $m$ times
behaved like a relativistic particle moving geodesically in an
effective 4-metric. Furthermore, this motion may also be described by
a classical motion where the strings are attracted
to the monopoles by a Newtonian inverse square force. Thus the
relativistic string motion in the one or two monopole case has all of
the symmetry of the classical Kepler and two gravitating centre problems.

Classical superstrings wound on the $x^4$ direction of (\ref{5met})
were considered in \cite{Gregory:1997te} in the context of Type II string theory
compactified on $T^6$. It was shown that since the surfaces of
constant $r$ correspond to squashed 3-spheres which are simply connected, it was possible to
unwind a string whilst keeping it arbitrarily far from a central
monopole. Since the winding number of a string is a conserved charge,
the H-electric charge, it was argued that the monopole should be able
to carry this charge and the zero mode which carries it was
identified.

An interesting question is whether there is an analogous behaviour for
winding strings in the LeBrun metrics. Since the LeBrun metrics are
not Ricci flat, it is not straightforward to exhibit a superstring
solution whose string metric is LeBrun. We shall therefore consider
the case of classical Nambu strings moving in a LeBrun background. We shall see that the string
motion may be thought of as particle motion in $\Ht$, with each
monopole attracting the string with a force given by the hyperbolic
generalisation of the Newtonian attraction.

We assume that we have a classical string moving in a background with
metric (\ref{5met}), where now $h_{ij}$ is the metric on $\Ht$ and $V$
takes instead the form
\begin{equation}
V = 1 + \frac{R_K}{2}\sum_{i=1}^k \left( \coth{D(P, P_i)}-1 \right),
\qquad d\omega= \star_h dV. \label{newtpot}
\end{equation}
The equations of motion are derived from the Nambu-Goto action, which may be
written
\begin{equation}
S=-\frac{1}{2\pi\alpha'}\int\left(\det-g_{AB}\pd{x^A}{u^a}\pd{x^B}{u^b} \right)^\frac{1}{2}du^1du^2.
\end{equation}
The motion of the string is specified by giving $x^A$ as a function of
the world-sheet coordinates $u^a=(\sigma, \tau)$. We shall consider a
closed string, so that the embedding functions must satisfy the
periodicity conditions
\begin{equation}
x^A(\sigma, \tau) = x^A(\sigma+2\pi, \tau).
\end{equation}
The simplest way to satisfy these equations is to consider a string
which winds $m$ times around the compact `internal' direction, $x^4$. The ansatz
we shall make is that the functions $x^A(\sigma, \tau)$ take the
form:
\begin{equation}
x^4 = m \sigma R_K, \qquad x^\alpha = x^\alpha(\tau), \label{ans}
\end{equation}
where $\alpha = 0, \ldots, 3$. By finding the equations of motion for
$x^A(\sigma, \tau)$ and
imposing the conditions (\ref{ans}) it is possible to show that the
equations of motion for $x^\alpha(\tau)$ take the form of geodesic
equations for a relativistic particle of mass $\mu = mR_k/\alpha'$ in an effective metric:
\begin{equation}
\tilde{g}_{\alpha \beta}dx^\alpha dx^\beta = -V^{-1}dt^2+h_{ij}dx^idx^j.
\end{equation}
Thus we may study the string motion by studying the Lagrangian
\begin{equation}
L = h_{ij}\dot{x}^i\dot{x}^j - V^{-1}\dot{t}^2
, \qquad \dot{} \equiv \pd{}{\tau}.
\end{equation}
We take $\tau$ to be an affine parameter so that $L=-\epsilon \mu^2$,
with $\epsilon = -1, 0, +1$ corresponding to spacelike, null and
timelike geodesics respectively. $t$ is clearly cyclic, so may be
eliminated in favour of its conjugate momentum $E$ and we arrive at
the energy conservation equation:
\begin{equation}
\frac{\mu^2}{2} h_{ij}\dif{x^i}{\tau}\dif{x^j}{\tau} + \frac{1}{2}(\epsilon - V)E^2 = \frac{\epsilon}{2}(E^2-\mu^2).
\end{equation}
Thus the string coordinates $x^\alpha$ behave like a particle of mass
$m$ moving
in $\Ht$ in a potential $(\epsilon - V)E^2/2$. From (\ref{newtpot}) it
is clear that this potential is that due to $k$ attractive particles,
whose force on the string is given by the hyperbolic Coulomb
interaction.

We have studied the case of one or two fixed Coulomb centres in $\Ht$
as special cases of the systems considered elsewhere. In particular, for
the one centre case, there is a conserved angular momentum and also a
conserved Runge-Lenz vector. In Beltrami coordinates, the path of the
string will be a conic section. Thus, if the string initially has some
angular momentum about the monopole, they can never collide. If the
string however starts with no angular momentum, it will eventually
reach the point where the $x^4$ dimension, and hence the string,  has
zero proper length. It seems plausible then that these monopoles
catalyse the annihilation of these winding strings, as was conjectured
in the case of Kaluza-Klein monopoles over flat space.

\section{Quantum Mechanics of the one-centre problem \label{quant}}

It was argued in \cite{Gibbons:1986df} that in a low energy regime
the quantum behaviour of two BPS monopoles may be approximated by
considering a wavefunction which obeys the Schr\"{o}dinger equation on
the two-monopole moduli space, whose metric is the Atiyah-Hitchin
metric. Since in this case the motion is geodesic, the Hamiltonian is
proportional to the covariant Laplacian constructed from the
Atiyah-Hitchin metric. If we are interested in quantum states where
the amplitude for the particles to be close together is small, we can
find approximate results by considering the self-dual Taub-NUT metric
as an approximation to the Atiyah-Hitchin metric for well separated
monopoles.

We are interested in the behaviour of well separated monopoles in hyperbolic
space. As was noted in section \ref{hypmon}, the absence of boost
symmetries means that the case of two particles interacting via a
Coulomb interaction does not reduce to a centre of mass
motion together with a fixed centre Coulomb problem
\cite{Maciejewski}. Thus, we expect that the problem of two monopoles
in hyperbolic space will require the solution of Schr\"{o}dinger's
equation on a $7$-dimensional manifold. In order to make analytic
progress, we instead consider the problem of the motion of a test
monopole in the field of one or more fixed monopoles. In the classical
case, we found that this can be interpreted as geodesic motion on a
manifold whose metric is the hyperbolic generalisation of the
multi-centre metrics. We found that the Lagrangian for geodesic motion
in the one-centre case admits an extra conserved vector of Runge-Lenz type, and we found the
most general potential which could be added such that this
persists. We shall solve the Schr\"{o}dinger equation for motion in
the one-centre metric with the potential found above. We shall find
the bound state energies and also the scattering amplitudes for a
particle scattered by the fixed centre. As special
cases, we have the case of two monopoles, and also the MIC-Kepler and
Kepler problems in hyperbolic space. We shall also show that the
Schr\"{o}dinger equation for the two centre case separates for pure
geodesic motion and with a potential generalising that of the
one-centre case.

\subsection{Bound states for the one-centre problem}

We wish to consider the quantum mechanics of a particle moving in a
space whose metric is given by:
\begin{equation}
ds^2 = V h + V^{-1} \left ( d \tau + \omega \right)^2. \label{metr1}
\end{equation}
With $h$ the metric on $\Ht$. The Laplace operator of this metric may be written in the form
\cite{Page:1979ga}:
\begin{equation}
\nabla^2 = \frac{1}{V} \widetilde{\Delta}_h + V
\frac{\partial^2}{\partial \tau^2}, \label{twistlap}
\end{equation}
where $\widetilde{\Delta}_h$ is the `twisted' Laplacian on $\Ht$ given
by
\begin{eqnarray}
\widetilde{\Delta}_h &=& h^{ij}\left(\nabla_i-\omega_i
\pd{}{\tau}\right)\left(\nabla_j-\omega_j \pd{}{\tau}\right) \nonumber
\\
 &=& \frac{(1-r^2)^2}{r^2}\left(\pd{}{r} r^2 \pd{}{r}\right) +
\frac{1-r^2}{r^2} \widetilde{\Delta}_{S^2},
\end{eqnarray} 
where for the second equality we take Beltrami coordinates on $\Ht$
and use the fact that $\omega_r=0$ to split the twisted Laplacian on
$\Ht$ into a radial part and a twisted Laplacian on $S^2$. The
time independent Schr\"{o}dinger equation for motion on the manifold with metric
(\ref{metr1}) in a potential $W$ is:
\begin{equation}
\left(-\frac{1}{2} \nabla^2 + W \right)\Psi = E\Psi.
\end{equation}
Using (\ref{twistlap}) this becomes:
\begin{equation}
\left(\widetilde{\Delta}_h + V^2 \pdd{}{\tau} - 2 WV + 2EV \right)\Psi
= 0.
\end{equation}

We wish to consider a Hamiltonian which classically admits a
Runge-Lenz vector, so we take
\begin{equation}
V(r) = k + \frac{\alpha}{r}, \qquad W(r) = \left (
\frac{\beta}{r}+\frac{e^2 p^2}{2 r^2} + \gamma \right )
\left(k+\frac{\alpha}{r}\right)^{-1}-\frac{1}{2}e^2
\left(k+\frac{\alpha}{r}\right).
\end{equation}
We will also impose the relation between $V$ and $\omega$ given by
(\ref{homega}), which implies that $\alpha=p$. In order to separate
variables for the Schr\"{o}dinger equation, we set $\tau = p\psi$ and
we make the following ansatz for a solution: 
\begin{equation}
\Psi_{KM}^J(r, \theta, \phi, \psi) = R_{KM}^J(r) D_{KM}^J(\theta,
\phi, \psi),
\end{equation}
where there is no sum implied over repeated indices. The functions
$D_{KM}^J$ are the Wigner functions. We replace the angular
derivatives by angular momentum operators according to
\begin{equation}
\widetilde{\Delta}_{S^2} = -(L_1^2 +L_2^2), \qquad \pdd{}{\psi}=-L_3^2.
\end{equation}
Acting on the Wigner functions we have that
\begin{equation}
L_1^2+L_2^2+L_3^2 = J(J+1), \qquad L_3^2=K^2=e^2p^2.
\end{equation}
Since $K$ is an integer or half-integer, we have a generalisation of the well known
result of Dirac that the existence of a magnetic monopole implies
charge quantisation once quantum mechanics is taken into account. We
also have the usual relation that $-J\leq K\leq J$. The
Schr\"{o}dinger equation now reduces to an ordinary differential
equation for the radial function:
\begin{equation}
\left[ \frac{\left(1-r^2 \right)^2}{r^2}\dif{}{r}\left( r^2 \dif{}{r} \right) -
\frac{A}{r^2}+\frac{B}{r} + C \right] R_{KM}^J(r) = 0\label{radeq}),
\end{equation}
where we have introduced some new constants
\begin{eqnarray}
A &=& J(J+1)\nonumber \\
B &=& 2Ep - 2\beta \nonumber \\
C &=& 2Ek - 2\gamma-p^2e^2+J(J+1).
\end{eqnarray}
Equation (\ref{radeq}) can be solved in terms of a hypergeometric
function. The solution regular at $r=0$ is:
\begin{equation}
R_{KM}^J(r) = \left(\frac{2r}{1+r}\right)^J
\left(\frac{1-r}{1+r}\right)^{\frac{1}{2}(1 + \sqrt{1+A-B-C})} F(a, b;
2J+2;\frac{2r}{1+r} ),
\end{equation}
where
\begin{eqnarray}
a &=&\frac{1}{2}(1+\sqrt{1+4A}+\sqrt{1+A-B-C}-\sqrt{1+A+B-C} \nonumber
\\b &=&\frac{1}{2}(1+\sqrt{1+4A}+\sqrt{1+A-B-C}+\sqrt{1+A+B-C}.
\end{eqnarray}
In order that the solution is also regular at $r=1$, we require that
$a$ is a non-positive integer. It is convenient to write $a=-N+J+1$.
This condition allows us to solve for $E$. Doing so, and replacing the
radius of the hyperbolic space $R_0$ by dimensional analysis we find that
\begin{equation}
E_N = \frac{p \beta -kN^2\pm N \sqrt{k^2N^2-2kp\beta +
    \frac{p^2}{R_0^2}(1-N^2+e^2p^2)+2\gamma p^2}}{p^2} \label{eng}
\end{equation}
is a bound state energy level, provided that $N$ takes one of the values $\abs{ep}+1,
\abs{ep}+2, \ldots$ and also that
\begin{equation}
-\beta + pE_N-N^2 \geq 0 \label{cond}
\end{equation}
is satisfied. We note a larger degeneracy than one might expect for
the spectrum of this problem. The energy depends only on the quantum
number $N$, not on the total angular momentum. This degeneracy is
exactly analogous to the degeneracy in the spectrum of the hydrogen
atom and is also due to the hidden symmetry whose classical
manifestation is the Runge-Lenz vector.

We may consider some of the special cases mentioned earlier. In the
limit of pure geodesic motion, which describes the interaction of a
hyperbolic monopole with a fixed monopole at the origin, we find that
the bound states have energy given by:
\begin{equation}
E_N = \frac{k}{p^2}\left(e^2p^2-N^2\pm
N\sqrt{N^2-e^2p^2+\frac{p^2}{k^2 R_0^2}(1-N^2+e^2p^2)} \right) \label{monlev}
\end{equation}
with $k=1+p/R_0$. Taking the limit $R_0 \to \infty$, we recover the
results for the energy levels of bound states of two Euclidean
monopoles as derived in \cite{Gibbons:1986df}. As in the Euclidean
case, we discard the states corresponding to the negative square root
in (\ref{monlev}) as these are tightly bound states which are not
described by our well separated monopole assumptions.

We can also recover the MICZ-Kepler problem on $\Ht$ as a limit of of our
system. This problem was studied by Kurochkin and Otchik
\cite{Kurochkin:2005} and by Meng in
other dimensions \cite{Meng:2005}. The limit we require is the limit
$p\to 0$, while keeping $ep=\mu$ fixed. This only makes sense in our expression for $E_N$
(\ref{eng}) if we take the positive sign for the square root. Taking
this limit, we find the spectrum for the MIC-Kepler problem in
hyperbolic space after setting $\gamma = -\mu^2/2$ to remove an
additive constant from the energy:
\begin{equation}
E_N = -\frac{\beta^2}{2N^2} - \frac{N^2-1}{2R_0^2}, \qquad N =
\abs{\mu}+1, \abs{\mu}+2, \ldots, \lfloor \sqrt{-\beta R_0} \rfloor,
\end{equation}
where $\lfloor x \rfloor$ is the largest integer not greater than
$x$. We may take the limit $R_0 \to \infty$ and we get the spectrum
for the MIC-Kepler problem in Euclidean space. Finally if we take
$R_0\to \infty$ and $\mu =0$ we have the spectrum of the hydrogen atom,
for some appropriate choice of units.

One interesting result of the above calculations is that the hydrogen
atom in hyperbolic space only has a finite number of bound state
energy levels, the number of levels being bounded above by
$\sqrt{-\beta R_0}$. A curious fact of the hydrogen atom in flat space
is that the partition function for a canonical ensemble of electrons
at temperature $T$ populating hydrogen atom energy levels does not converge:
\begin{equation}
Z \stackrel{\text{\tiny !}}{=} \sum_{N=1}^\infty e^{\beta^2/TN^2}.
\end{equation}
If the hydrogen atom is in hyperbolic space however, the partition
function converges since the sum is cut off at some finite value of
$N$. This is reminiscent of large black holes in AdS, which may be thought
of as having been put in a `box' in order to stabilise them against
losing their energy through Hawking radiation.

\subsection{Schr\"{o}dinger's equation in pseudoparabolic coordinates}

When considering Rutherford scattering in Euclidean space, i.e.\ the
problem of an alpha particle scattered by a much heavier nucleus, it
is convenient to work with parabolic coordinates, $\xi, \eta$. These
are defined in terms of the usual radial coordinate $r$ and the $z$
coordinate by
\begin{equation}
\xi = r+z, \qquad \eta = r-z.
\end{equation}
In these coordinates the Schr\"{o}dinger equation separates and it is
possible to solve the resulting ODEs to find the Rutherford scattering
formula \cite{Landau:1959}. 

In hyperbolic space, the appropriate choice of coordinates for this
scattering problem are the pseudoparabolic coordinates defined in section
\ref{ppara}. As in the previous section, we have that
Schr\"{o}dinger's equation takes the form:
\begin{equation}
\left(\widetilde{\Delta}_h + V^2 \pdd{}{\tau} - 2 WV + 2EV \right)\Psi
= 0 \label{sch}
\end{equation}
with
\begin{equation}
\widetilde{\Delta}_h = h^{ij}\left(\nabla_i-\omega_i
\pd{}{\tau}\right)\left(\nabla_j-\omega_j \pd{}{\tau}\right).
\end{equation}
Now the coordinates on hyperbolic space are $\mu, \nu$ as defined in
section \ref{ppara}. A very similar calculation to that performed in
section \ref{actwav} to separate variables for the Hamilton-Jacobi
equation leads to separation of variables for the Schr\"{o}dinger
equation. More precisely, we can write
\begin{equation}
\Psi(\mu, \nu, \phi, \tau) = e^{i m \phi}e^{ie\tau}M(\mu)N(\nu)
\end{equation}
and the Schr\"{o}dinger equation (\ref{sch}) implies the two second
order ordinary differential equations
\begin{eqnarray}
\frac{1+\mu^2}{\mu}\dif{}{\mu}\left(\mu\dif{M}{\mu} \right) -
\frac{(m+ep)^2}{\mu^2}M -
\frac{2E(k-p)+2\beta-2\gamma-e^2p^2}{1+\mu^2}M &=& KM \nonumber \\
\frac{1-\nu^2}{\nu}\dif{}{\nu}\left(\nu \dif{N}{\nu} \right) - 
\frac{(m-ep)^2}{\nu^2}N +
\frac{2E(k+p)-2\beta-2\gamma-e^2p^2}{1-\nu^2} N&=&-KM,
\end{eqnarray}
$K$ is here a separation constant.

These equations may be put into the form of the hypergeometric
equation and solved. The solution for $M(\mu)$, regular at $\mu=0$, is
given in terms of the hypergeometric function by
\begin{equation}
M(\mu) = \mu^{\abs{m+ep}}(1+\mu^2)^{-\frac{1}{2}\abs{m+ep} -
  \frac{1}{2}\sqrt{K}}F\left( a_1, b_1; c_1; \frac{\mu^2}{1+\mu^2}
  \right). \label{solM}
\end{equation}
New constants have been defined:
\begin{eqnarray}
a_1 &=&
\frac{1}{2}\abs{m+ep}+\frac{1}{2}\sqrt{K}
+\frac{1}{2}+\frac{1}{2}\sqrt{1-A}
\nonumber \\
b_1 &=&
\frac{1}{2}\abs{m+ep}+\frac{1}{2}\sqrt{K}
+\frac{1}{2}-\frac{1}{2}\sqrt{1-A}
\nonumber \\
c_1 &=& 1+\abs{m+ep} \nonumber \\
A&=& 2E(k-p)+2\beta-2\gamma-e^2p^2.
\end{eqnarray}
Similarly, the solution for $N(\nu)$ regular at $\nu=0$ is given by
\begin{equation}
N(\nu) = \nu^{\abs{m-ep}}(1-\nu^2)^{\frac{1}{2}+\frac{1}{2}\sqrt{1-B}}
F\left(a_2, b_2; c_2; \nu^2 \right) \label{solN}
\end{equation}
with
\begin{eqnarray}
a_2 &=&\frac{1}{2}\abs{m-ep}+\frac{1}{2}+\frac{1}{2}\sqrt{1-B} -
\frac{\sqrt{K}}{2} \nonumber \\ b_2 &=&\frac{1}{2}\abs{m-ep}+\frac{1}{2}+\frac{1}{2}\sqrt{1-B} +
\frac{\sqrt{K}}{2} \nonumber \\ c_2 &=& 1+\abs{m-ep} \nonumber \\ B &=& 2E(k+p)-2\beta-2\gamma-e^2p^2.
\end{eqnarray}
We can recover the bound state energy levels from these formulae by
imposing that the wavefunction is also regular at $\mu =\infty$ and
$\nu=1$. This will occur if:
\begin{equation}
a_1 = -n_1, \qquad a_2 = -n_2
\end{equation}
where $n_1, n_2$ are integers called the parabolic quantum
numbers. This gives exactly the same energies for the bound states as
in the radial calculation and a counting of states shows that they
have the same degeneracy. 

\subsection{Scattering States}

We are more interested in the scattering states. In flat space, we
seek to write the wavefunction at large $r$ as the sum of an incoming
plane wave and an outgoing scattered spherical wave:
\begin{equation}
\Psi \sim e^{-ikz} + f(\theta) \frac{e^{ikr}}{r},
\end{equation}
we then interpret $f(\theta)$ as the amplitude per unit solid angle for a
particle to be scattered at angle $\theta$ to the incoming wave. In
hyperbolic space, the situation is very similar. Instead of plane
waves, the appropriate objects in $\Ht$ are horospherical waves which
are constant on the horospheres (see section \ref{horo}). To discuss
the scattering solutions of Schr\"{o}dinger's equation, it is convenient to use the coordinates $(\chi,
\theta, \phi)$ for $\Ht$, in which the metric takes the form:
\begin{equation}
h = d\chi^2 + \sinh^2 \chi\left(d\theta^2+\sin^2\theta d\phi^2 \right).
\end{equation}
In these coordinates, the horospherical wave solution of
Schr\"{o}dinger's equation in hyperbolic space with no potential is given by:
\begin{equation}
\Psi = (\cosh \chi - \sinh \chi \cos \theta)^{-1-i\kappa'} \label{schhoro}
\end{equation}
and the spherical wave solution is:
\begin{equation}
\Psi = \frac{e^{i\kappa'\chi}}{\sinh{ \chi}},
\end{equation}
where $\kappa'$ is related to the energy $E$ by $\kappa' =
\sqrt{2E-1}$. Thus we shall seek a solution to the full
Schr\"{o}dinger equation (\ref{sch}) which has the asymptotic form as
$\chi \to \infty$
\begin{equation}
\Psi \sim  (\cosh \chi - \sinh \chi \cos \theta)^{-1-i\kappa'} +
f(\theta)\frac{e^{i\kappa'\chi}}{\sinh{ \chi}} \label{asy}
\end{equation}
and we will again interpret $f(\theta)$ as a scattering amplitude. In
order to find a solution with these asymptotics, we look once again at
the solution in pseudoparabolic coordinates. It is straightforward to
show that (\ref{schhoro}) in pseudoparabolic coordinates becomes
\begin{equation}
\Psi=\left[(1+\mu^2)(1-\nu^2) \right]^{\frac{1}{2}+\frac{i\kappa'}{2}}.
\end{equation}
The condition on the asymptotic form of the wavefunction as $\chi \to
\infty$ becomes a condition as $\nu \to 1$. We define $\kappa$ and
$\lambda$ to be
\begin{equation}
\kappa = \sqrt{2E(p+k)-1-2\gamma-2\beta+e^2p^2}, \qquad \lambda = \sqrt{2E(k-p)-1-2\gamma+2\beta-e^2p^2}.
\end{equation}
In order to get the correct asymptotics, we set $m=-ep$ and
$K=-1-i\lambda$. The solutions (\ref{solM}, \ref{solN}) then simplify
and we get:
\begin{eqnarray}
M(\mu) &=& (1+\mu^2)^{\frac{1}{2}+\frac{i\lambda}{2}} \nonumber \\
N(\nu) &=& \nu^{2
  \abs{ep}}(1-\nu^2)^{\frac{1}{2}+\frac{i\kappa}{2}}F\left(\abs{ep}+1+\frac{i}{2}(\kappa+\lambda), \abs{ep}+\frac{i}{2}(\kappa-\lambda); 1+2\abs{ep};\nu^2 \right)
\end{eqnarray}
As $\nu \to 1$, we can expand the hypergeometic function using the
formula:
\begin{equation}
F(a,b;c;x) \stackrel{\text{\scriptsize $x \to 1$}}{\sim}
\frac{\Gamma(c)\Gamma(c-a-b)}{\Gamma(c-a)\Gamma(c-b)} + (1-x)^{c-a-b} \frac{\Gamma(c)\Gamma(a+b-c)}{\Gamma(a)\Gamma(b)}.
\end{equation}
Performing the expansion and multiplying by a factor to normalise the
incoming wave, we find
\begin{equation}
\Psi  \stackrel{\text{\scriptsize $\nu \to 1$}}{\sim}
\left[(1+\mu^2)(1-\nu^2)
  \right]^{\frac{1}{2}+\frac{i\kappa}{2}}(1+\mu^2)^{\frac{i}{2}(\lambda-\kappa)} +
\Delta (1-\nu^2)^{\frac{1}{2}-\frac{i\kappa}{2}}(1+\mu^2)^{\frac{1}{2}+\frac{i\lambda}{2}},
\end{equation}
where
\begin{equation}
\Delta = \frac{\Gamma(i\kappa) \Ga{\abs{ep}-\frac{i}{2}(\kappa+\lambda)}
  \Ga{\abs{ep}+1-\frac{i}{2}(\kappa-\lambda)}}{\Ga{-i\kappa}
  \Ga{\abs{ep}+1+\frac{i}{2}(\kappa+\lambda)} \Ga{\abs{ep}+\frac{i}{2}(\kappa-\lambda)}}
\end{equation}
We can now return to the $(\chi,\theta, \phi)$ coordinates, and we
find that this formula becomes:
\begin{equation}
\Psi  \stackrel{\text{\scriptsize $\chi \to \infty$}}{\sim} (\cosh
\chi - \sinh \chi \cos
\theta)^{-1-i\kappa}e^{-i(\lambda-\kappa)\ln\sin\frac{\theta}{2}} +
\frac{\Delta}{2}\mathrm{cosec}^2 \frac{\theta}{2}
e^{-i(\lambda-\kappa)\ln\sin\frac{\theta}{2}}
\frac{e^{i\kappa\chi}}{\sinh{ \chi}}
\end{equation}
We thus find the scattering amplitude to be:
\begin{equation}
f(\theta) = \frac{\Delta}{2} \mathrm{cosec}^2 \frac{\theta}{2}.
\end{equation}
It is interesting to note that both the incoming and outgoing wave are
distorted relative to the proposed asymptotic form (\ref{asy}). This
occurs also in the Euclidean case, because the Coulomb potential does
not decay fast enough to justify an expansion in terms of free waves
far from the scattering centre.

It is possible to recover the bound state energy levels by making the
analytic continuations $\kappa \to i \tilde{\kappa}$ and $\lambda \to
i \tilde{\lambda}$. We then find that the scattering amplitude has
poles at precisely the values of $E$ which were found previously to be
bound state energies. Another useful check of this formula is to use it to find the Rutherford
scattering formula. We first set $p=\gamma=e=0$ and $k=\beta=1$, then
we replace the hyperbolic radius $R_0$ by dimensional
analysis. Finally we let $R_0 \to \infty$ and we find
\begin{equation}
f(\theta) = - \frac{1}{2k'^2\sin^2\frac{\theta}{2}} \frac{\Ga{1+i/k'}}{\Ga{1-i/k'}}
\end{equation}
where $E = \frac{1}{2}k'^2$. This is precisely Rutherford's formula, as
found in \cite{Landau:1959}.

\section{The two centre problem \label{twocent}}

\subsection{Separability of Schr\"{o}dinger's equation for the
  two-centre problem}

In \cite{Gibbons:1987sp} it was shown that the Schr\"{o}dinger
equation for the two-centre Taub-NUT metric separates in spheroidal
coordinates. This generalises the fact that the Schr\"{o}dinger
equation for a diatomic molecule separates in the same way that the
Runge-Lenz vector for Taub-NUT generalises that of the Hydrogen
atom. We find here that in a suitable set of coordinates, the
two-centre LeBrun metric has a separable Schr\"{o}dinger equation. The
coordinates are the pseudospheroidal coordinates used by Vozmischeva
\cite{vosm}, which for convenience are also described in section (\ref{pssp})

We consider motion in a space with metric
\begin{equation}
ds^2 = V h + V^{-1} (d \tau +\omega)^2 \label{dmet}
\end{equation}
Where $V$ and $\omega$ are given by
\begin{equation}
V(P) = k-p_1 \coth D(P,P_1)-p_2\coth D(P, P_2), \qquad d\omega =
\star_hdV. \label{defA}
\end{equation}
It is convenient to work for the moment with the pseudosphere model of
Hyperbolic space (see section \ref{pseudosphere}). The general point $P \in \Ht$ has coordinates $(W,
X, Y, Z) \in \mathbb{E}^{3,1}$ which are subject to the constraint $W^2-X^2-Y^2-Z^2=1$. We
suppose that the fixed centres are at $(\alpha, 0,0, \pm\beta)$. The
distance between two points $P, P'$ is given in terms of the
$SO(3,1)$ invariant inner product on $\mathbb{E}^{3,1}$ by:
\begin{equation}
D(P, P') = \theta, \qquad \mathrm{where} \qquad \cosh \theta =
\frac{\langle P, P' \rangle}{\norm{P}\norm{P'}} = WW' -
XX'- YY' - ZZ'.
\end{equation}
Thus the potential function $V$ takes the form in these coordinates
\begin{equation}
V = k-p_1\coth{\theta_+}-p_2\coth{\theta_-}, \qquad \mathrm{with}\qquad
\cosh \theta_{\pm} =  \alpha W \mp \beta Z. \label{doubleV}
\end{equation}
In order to pass to the pseudospheroidal coordinates, we first take
polar coordinates $(P, \phi)$ in the $X$-$Y$ plane. The
pseudospheroidal coordinates $\xi$, $\eta$ are then defined in terms of
$W, P, Z$ by the relations (\ref{psdspher}):
\begin{eqnarray}
\nonumber W &=& \frac{1}{\alpha}\sqrt{(\alpha^2-\xi^2)(\alpha^2+\eta^2)},
\\
\nonumber Z &=&
\frac{\mathrm{sgn}(Z)}{\beta}\sqrt{(\beta^2-\xi^2)(\beta^2+\eta^2)},
  \\
P &=& \frac{\xi \eta}{\alpha \beta}.
\end{eqnarray}
The metric on $\Ht$ in these coordinates is given by (\ref{spheroidalmetric})
\begin{equation}
ds^2 = \frac{\xi^2+\eta^2}{(\alpha^2-\xi^2)(\beta^2-\xi^2)} d\xi^2 +
\frac{\xi^2+\eta^2}{(\alpha^2+\eta^2)(\beta^2+\eta^2)}d\eta^2 +
\frac{\xi^2\eta^2}{\alpha^2\beta^2}d\phi^2 \label{sphmet}
\end{equation}
Using these relations with (\ref{doubleV}) it is possible after some
algebra to find the following formula for $V$ in terms of the
pseudospheroidal coordinates:
\begin{equation}
V = k - (p_1+p_2)
\frac{\sqrt{(\alpha^2+\eta^2)(\beta^2+\eta^2)}}{\eta^2+\xi^2} -
  \mathrm{sgn}(Z)(p_1-p_2)\frac{\sqrt{(\alpha^2-\xi^2)(\beta^2-\xi^2)}}{\eta^2+\xi^2}.
\end{equation}
The simplest way to compute $\omega$ is directly from the formula in
(\ref{defA}). Using (\ref{sphmet}), again after some algebra, it can
be shown that $\omega$ takes the form:
\begin{equation}
\omega = \frac{1}{\alpha\beta} \left ( (p_1+p_2)\eta^2
\frac{\sqrt{(\alpha^2-\xi^2)(\beta^2-\xi^2)}}{\eta^2+\xi^2}   -
\mathrm{sgn}(Z)(p_1-p_2)\xi^2 \frac{\sqrt{(\alpha^2+\eta^2)(\beta^2+\eta^2)}}{\eta^2+\xi^2} \right ) d \phi.
\end{equation}

We consider the Schr\"{o}dinger equation for the metric (\ref{dmet})
with a potential $W$. The Schr\"{o}dinger equation can be re-written
in the form (\ref{sch}). We make a separation ansatz
\begin{equation}
\Psi_{m e}(\xi, \eta, \tau, \phi) = e^{i m \phi}e^{i e \tau}X(\xi)Y(\eta)
\end{equation}
and we find that the Schr\"{o}dinger equation separates when $W$
satisfies
\begin{equation}
WV = \gamma +\beta_+ \coth \theta_+ +\beta_- \coth \theta_-.
\end{equation}
In this case, the functions $X(\xi)$ and $Y(\eta)$ obey the ordinary
differential equations
\begin{eqnarray}
KX &=&\frac{\sqrt{(\alpha^2-\xi^2)(\beta^2-\xi^2)}}{\xi}\dif{}{\xi}\left(
\xi \sqrt{(\alpha^2-\xi^2)(\beta^2-\xi^2)} \dif{X}{\xi} \right) +
\left(A^+\xi^2-\frac{B^+}{\xi^2} \right) X \nonumber\\ && +
\left(C^+\mathrm{sgn}(Z)+\frac{D^+}{\xi^2}\right)\sqrt{(\alpha^2-\xi^2)(\beta^2-\xi^2)}
X 
\end{eqnarray}
and 
\begin{eqnarray}
-KY &=&\frac{\sqrt{(\alpha^2+\eta^2)(\beta^2+\eta^2)}}{\eta}\dif{}{\eta}\left(
\xi \sqrt{(\alpha^2+\eta^2)(\beta^2+\eta^2)} \dif{Y}{\eta} \right) +
\left(A^-\eta^2-\frac{B^-}{\eta^2} \right) Y \nonumber\\ && +
\left(C^-+\mathrm{sgn}(Z)\frac{D^-}{\eta^2}\right)\sqrt{(\alpha^2+\eta^2)(\beta^2+\eta^2)}
Y.
\end{eqnarray}
$K$ is a separation constant. In the case where the potential
vanishes, $K$ corresponds to a Killing tensor of the metric
(\ref{dmet}). We have defined some new constants
\begin{eqnarray}
A^{\pm} &=& 2Ek-2\gamma-e^2k^2-e^2(p_1\mp p_2)^2, \nonumber \\
B^{\pm} &=& \alpha^2\beta^2(m^2+e^2(p_1\pm p_2)^2), \nonumber \\
C^\pm &=& 2e^2(p_1\mp p_2)k - 2E(p_1\pm p_2)-2(\beta_1\mp \beta_2),
\nonumber \\
D^\pm &=& \pm 2em\alpha\beta(p_1\pm p_2).
\end{eqnarray}

\subsection{Separability of the Hamilton-Jacobi equation for the
  two-centre problem}

Additive separability of the Hamilton-Jacobi equation follows, formally
at least, from
the multiplicative separability of the Schr\"{o}dinger equation using
a semi-classical approximation. It is however straightforward, using
the expressions for $V$ and $\omega$ to separate variables for the
Hamilton-Jacobi equation in pseudospheroidal coordinates
explicitly. We find that the Hamilton-Jacobi function takes the form:
\begin{equation}
S = \phi p_\phi + \tau p_\tau + M(\xi) + N(\eta),
\end{equation}
where $M$, $N$ satisfy ordinary differential equations:
\begin{eqnarray}
\tilde{K} &=& (\alpha^2-\xi^2)(\beta^2-\xi^2)\left(\dif{M}{\xi}
\right)^2  + a^+\xi^2+\frac{b^+}{\xi^2} \nonumber \\ && +
\left(c^+\mathrm{sgn}(Z)+\frac{d^+}{\xi^2}
\right)\sqrt{(\alpha^2-\xi^2)(\beta^2-\xi^2)}
\end{eqnarray}
and
\begin{eqnarray}
-\tilde{K} &=& (\alpha^2+\eta^2)(\beta^2+\eta^2)\left(\dif{N}{\eta}
\right)^2  + a^-\eta^2+\frac{b^-}{\eta^2} \nonumber \\ && +
\left(c^-+\mathrm{sgn}(Z)\frac{d^-}{\eta^2}
\right)\sqrt{(\alpha^2+\eta^2)(\beta^2+\eta^2)}.
\end{eqnarray}
The new constants are
\begin{eqnarray}
a^{\pm} &=& p_\tau{}^2+p_\tau{}^2(p_1\mp p_2)^2+2\gamma-2Ek^2, \nonumber
\\
b^\pm &=&\alpha^2\beta^2(p_\phi{}^2+(p_1\pm p_2)^2p_\tau{}^2),
\nonumber \\
c^\pm &=& 2E(p_1 \mp p_2)+2(\beta_1\mp\beta_2)-2kp_\tau{}^2(p_1\mp
p_2), \nonumber \\
d^\pm &=& \pm 2 \alpha \beta p_\phi p_\tau (p_1\pm p_2).
\end{eqnarray}

Thus the two centre problem, with a potential generalised from that of
the one centre problem, is classically integrable in the Liouville
sense, i.e.\ there are sufficiently many conserved quantities to
reduce the problem to quadratures. The `extra' conserved quantity,
$\tilde{K}$ corresponds in the absence of a potential to a rank 2
Killing tensor of the metric (\ref{dmet}). 

The system is also quantum mechanically integrable. In
the operator formalism, this corresponds to being able to find a basis
of states which are eigenstates of operators which
commute with the quantum Hamiltonian. This quantum integrability is a
stronger result than classical integrability. In \cite{Carter:1977pq}
Carter showed that a classical rank 2 Killing tensor will not in general generate a
conserved charge in the quantum regime. There is an anomaly given by:
\begin{equation}
\com{\hat{K}}{H} \propto \left(K^{\nu[\mu}R^{\sigma]}{}_\nu
\right)_{;\sigma} \hat{p}_\mu.
\end{equation}
For a Ricci flat background, this anomaly will always vanish and there
will be a conserved quantum charge associated to the classical
conserved quantity. In our case the Ricci tensor does not vanish so
{\it a priori} we cannot expect a classical symmetry to give rise to a
quantum conserved charge. The separation of variables for the
Schr\"{o}dinger equation shows however that the anomaly vanishes in
this case.

\section{Conclusion}

We have investigated the motion of hyperbolic monopoles in the limit
where the monopoles are well separated and found many similarities
with monopoles in Euclidean space. Because of the absence of
boost symmetries for hyperbolic space it was necessary to consider
the case where $N$ monopoles are fixed while one is free to move
in the asymptotic field of the others. We found that this motion may
be interpreted as a geodesic motion on a space whose metric is a
LeBrun metric of negative mass. We interpret this as the metric on a
non-geodesic surface contained within the full moduli space of the
theory.

We then investigated in depth the classical and quantum motion of a
monopole in the presence of a single fixed monopole at the origin. It
was found that this system may be considered within a broader class of
systems which have similar properties arising from the existence of an
extra `hidden' constant of the motion which generalises the Runge-Lenz
vector of the classical Kepler problem. These systems have an enhanced
dynamical symmetry algebra and this results in some attractive
properties. The system is integrable in the Liouville sense and one
can show that the orbits, when expressed in suitable coordinates, are
conic sections. The hodograph, suitably defined, is elliptic and
becomes circular precisely when the system under consideration is the
hyperbolic Kepler system. The Hamilton-Jacobi equation separates in
both spherically symmetric coordinates and pseudoparabolic coordinates,
enabling the surfaces of constant action for scattering particles to
be found analytically. 

The classical integrability persists in the
quantum regime and the Schr\"{o}dinger equation separates in
spherically symmetric and pseudoparabolic coordinates. This enables us
to calculate the bound state energies and the scattering amplitude for
a monopole moving in the field of another fixed at the origin. We find
similarities with the hydrogen atom problem: the degeneracy of the
energy levels is enhanced by the existence of the Runge-Lenz vector;
the scattering amplitudes have the same angular dependence as appears
in the Rutherford formula.

We have also considered the problem of a monopole moving in the field
of two fixed monopoles and we find that it is integrable both
as a classical and a quantum system. Geometrically the integrability
properties of both the one and two centre cases are due to the
existence of rank 2 Killing tensors. These tensors generate
non-geometric transformations of the phase space which are symmetries
of the system and give rise to conserved quantities which are
quadratic in the momenta.

Thus we have found many similarities between the motion of a hyperbolic
monopole about a fixed monopole and the same problem in flat
space. The systems we considered generalise the original Kepler
problem and exhibit many of beautiful properties which arise as a
consequence of its symmetries.
\\

{\large {\bf Acknowledgements\\}}

CMW would like to thank Sean Hartnoll, Julian Sonner and especially
Gustav Holzegel for discussions and to acknowledge funding from PPARC.

\newpage
\appendix
\section{Some useful properties of Hyperbolic Space}
\renewcommand{\theequation}{A-\arabic{equation}}
\setcounter{equation}{0}  

\subsection{Coordinates on Hyperbolic space} 
For convenience, we collect here several coordinate systems for
hyperbolic space which will prove useful in this work. 

\subsubsection{The pseudosphere model}
\label{pseudosphere}

The most easily 
visualised model of hyperbolic space is as the upper leaf of the unit
pseudosphere in Minkowski space. This model shows most clearly the
parallels between spherical and hyperbolic geometry:
\begin{equation}
\label{psdsphr} 
\Ht = \left \{ (W, X, Y, Z) \in \mathbb{E}^{3,1} : W^2-X^2-Y^2-Z^2=1,
W>0 \right \} 
\end{equation}
where the metric on $\mathbb{E}^{3,1}$ is given by
\begin{equation}
\label{minkowski}
ds^2=-dW^2+dX^2+dY^2+dZ^2.
\end{equation}
The metric on $\Ht$ is then the restriction of this metric to the
tangent space of the pseudosphere. Lorentz transformations of the full
space $\mathbb{E}^{3,1}$ preserve both the metric (\ref{minkowski})
and the pseudosphere condition (\ref{psdsphr}) and so correspond to
isometries of $\Ht$. We see that $SO(3,1)$ acts transitively on $\Ht$,
and we may write $\Ht = SO(3,1)/SO(3)$. As in the case of the three sphere in
$\mathbb{E}^4$, the geodesics are given by the
intersections of 2-planes through the origin with the
pseudosphere. It is useful to define the various coordinate systems
that we will use in terms of this model.

\subsubsection{Poincar\'{e} Coordinates}
\label{confflat}

Among the best known of the coordinate systems on $\Ht$ are the
Poincar\'{e} coordinates. These are obtained from the pseudosphere
model by the analogue of stereographic projection.

\begin{figure}[!ht]
\[
\input{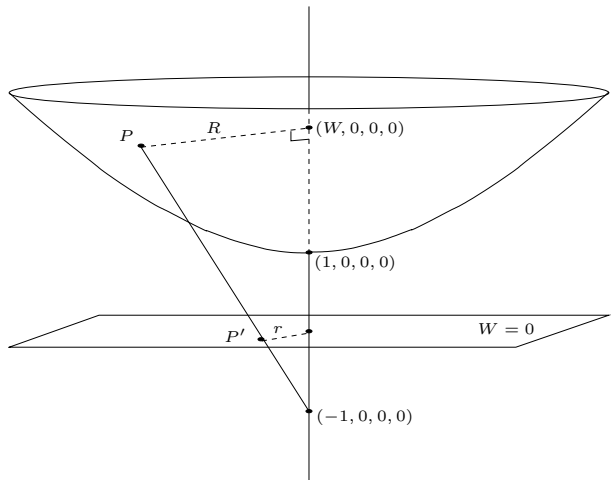}
\]
\caption{Stereographic projection of $\Ht$ onto plane through the
  origin}
\label{conf}
\end{figure}

We consider the projection through the point $(-1,0,0,0)$ onto the plane with normal $(1, 0, 0, 0)$
in $\mathbb{E}^{3,1}$ which passes through the origin. This situation
is shown with one dimension suppressed in Figure \ref{conf}. From the
diagram, we see that the point $P$ in the pseudosphere is mapped to a
point $P'$ on the plane $W=0$. We introduce cylindrical polar
coordinates for $\mathbb{E}^{3,1}$ according to:
\begin{eqnarray}
\nonumber X &=& R \sin \Theta \sin \Phi \\
\nonumber Y &=& R \sin \Theta \cos \Phi \\
Z &=& R \cos \Theta.
\label{cylind}
\end{eqnarray}
We denote the coordinates of $P'$ by $(0, r,
\theta, \phi)$. Clearly from the diagram we have that the coordinates
of $P'$ are related to those of $P$ by
\begin{equation}
\frac{R}{W+1} = \frac{r}{1}, \qquad \quad \Theta = \theta, \qquad \quad \Phi = \phi.
\end{equation}
Where the condition that $P$ lie on the pseudosphere becomes
$W^2-R^2=1$. We note that since $W>R$, we must have $r<1$, so $\Ht$
has been mapped to the interior of the unit ball in
$\mathbb{R}^3$. The sphere $r=1$ corresponds to the `sphere at
infinity' of $\Ht$. In the $r$, $\theta$, $\phi$ coordinates the metric on $\Ht$ can
be written as
\begin{equation}
\label{conformalmetric}
ds^2 = \frac{4}{(1-r^2)^2}\left \{ dr^2 + r^2(d \theta^2 + \sin^2\theta
d \phi^2) \right \}.
\end{equation}
This is the model of $\Ht$ known as the Poincar\'{e} ball. We
recognise the term in braces as the flat metric on $\mathbb{R}^3$ so
that the metric is conformally flat in these coordinates. The
geodesics in this model correspond to the arcs of circles which
intersect the sphere $r=1$ orthogonally.

\subsubsection{Beltrami Coordinates}
\label{beltrcoords}
\begin{figure}[!ht]
\[
\input{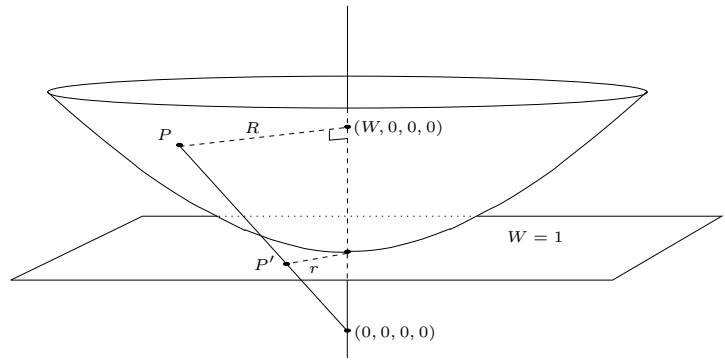}
\]
\caption{Gnomonic projection of $\Ht$ onto the tangent plane at $(1,0,0,0)$}
\label{belt}
\end{figure}

The Beltrami coordinates on $\Ht$ are the hyperbolic analogue of the
gnomonic coordinates on $S^3$. They are again defined by a projection,
but this time we project through the origin onto the tangent plane to
$\Ht$ at $(1,0,0,0)$, as shown in Figure \ref{belt}

We can again write the coordinates of $P'$ in terms of those of
$P$. We note that the projection will change the magnitude, but not
the direction of the
vector $\mathbf{X} = (X,Y,Z)$. This is simply the observation that the
angles $\Phi$ and $\Theta$ are unchanged by the projection. The
Beltrami coordinates $\mathbf{r}$ are defined in terms of the
coordinates in $\mathbb{E}^{3,1}$ according to:
\begin{equation}
\mathbf{r}=\frac{\mathbf{X}}{W} \label{beltcoord}
\end{equation}
Once again, the condition that $P$ lies on the pseudosphere is
$W^2-\vect{X}^2=1$, so we are again mapping $\Ht$ to the interior of the
unit ball. The metric in Beltrami coordinates is given by
\begin{equation}
\label{beltramimetric1}
ds^2= \frac{d \vect{r}^2}{1-\abs{\vect{r}}^2} + \frac{(\vect{r}\cdot d
  \vect{r})^2}{\left ( 1-\abs{\vect{r}}^2 \right )^2}.
\end{equation}
Defining spherical polar coordinates on $\mathbb{R}^3$ in the usual
way, the metric may alternatively be written
\begin{equation}
\label{beltramimetric2}
ds^2= \frac{dr^2}{\left ( 1-r^2 \right )^2} + \frac{r^2}{ 1-r^2} (d
\theta^2 + \sin^2 \theta d \phi^2)
\end{equation}
Since the geodesics on $\Ht$ are given by the intersection
with 2-planes through the origin in the pseudosphere model,
it is straightforward to see that they project back to straight lines in
the Beltrami ball model.

\subsubsection{Pseudospheroidal Coordinates \label{pssp}}

When considering the two centre problem in flat space, it is
convenient to work in spheroidal coordinates. Since the two centres
may be taken to lie at $(0, 0, \pm a)$, we take the polar angle $\phi$ as
one of the coordinates. The other two coordinates are defined in terms
of the distances to the centres, $r_+$ and $r_-$ by 
\begin{equation}
\zeta = \frac{r_+ + r_-}{2a}, \qquad \lambda = \frac{r_+ - r_-}{2a}.
\end{equation}
The coordinate surfaces $\zeta = \mathrm{const.}$ are prolate
spheroids and the surfaces  $\lambda = \mathrm{const.}$ are
hyperboloids. In these coordinates the Laplacian on flat space
separates.

In our investigations we shall require an analogous set of coordinates
for $\Ht$. We shall use the coordinates defined by Vozmischeva in \cite{vosm}. We first set
\begin{equation}
X=P \cos \phi, \qquad Y = P \sin \phi,
\end{equation}
so that the metric on $\mathbb{E}^{3,1}$ becomes
\begin{equation}
ds^2 = -dW^2+dZ^2+dP^2+P^2 d \phi^2 \label{cylpolmet}
\end{equation}
and the equation of the pseudosphere becomes $W^2-P^2-Z^2=1$. We assume that we have two centres located at $(\alpha,0,0, \pm \beta)$, where of course $\alpha^2-\beta^2=1$. The pseudospheroidal
coordinates $\xi, \eta$ are defined by
\begin{eqnarray}
\nonumber W &=& \frac{1}{\alpha}\sqrt{(\alpha^2-\xi^2)(\alpha^2+\eta^2)},
\\
\nonumber Z &=&
\frac{\mathrm{sgn}(Z)}{\beta}\sqrt{(\beta^2-\xi^2)(\beta^2+\eta^2)},
  \\
P &=& \frac{\xi \eta}{\alpha \beta},  \label{psdspher}
\end{eqnarray}
with
\begin{equation}
0 \leq \xi \leq \beta, \qquad 0 \leq \eta < \infty
\end{equation}
Since $\mathrm{sgn}(Z)$ enters these equations, the equations
(\ref{psdspher}) in fact define two coordinate patches, which together
cover $\Ht$ and overlap along the plane $Z=0$. From these equations, we can deduce the relations
\begin{equation}
\frac{W^2}{\alpha^2-\xi^2} =
\frac{Z^2}{\beta^2-\xi^2}-\frac{P^2}{\xi^2}, \qquad \frac{W^2}{\alpha^2+\eta^2} =
\frac{Z^2}{\beta^2+\eta^2}+\frac{P^2}{\eta^2} \label{cordsurf}
\end{equation}
In order to interpret the geometric meaning of the $\xi, \eta$
coordinates, we may use the gnomonic projection and consider the
coordinate surfaces in Beltrami coordinates. Using cylindrical polars
on the Beltrami ball, from above we have that
\begin{equation}
z = \frac{Z}{W}, \qquad \rho = \frac{P}{W},
\end{equation}
while $\phi$ remains unchanged. Thus, in Beltrami coordinates, the
equations which determine the coordinate surfaces of $\xi$ and $\eta$
become:
\begin{eqnarray}
\frac{1}{\alpha^2-\xi^2} &=& \frac{z^2}{\beta^2-\xi^2} -
\frac{\rho^2}{\xi^2} \label{hyperb}\\
\frac{1}{\alpha^2+\eta^2} &=&
\frac{z^2}{\beta^2+\eta^2}+\frac{\rho^2}{\eta^2} \label{ellip}.
\end{eqnarray}

\begin{figure}[!ht]
\centering \framebox {\includegraphics[height=3in,
width=3in]{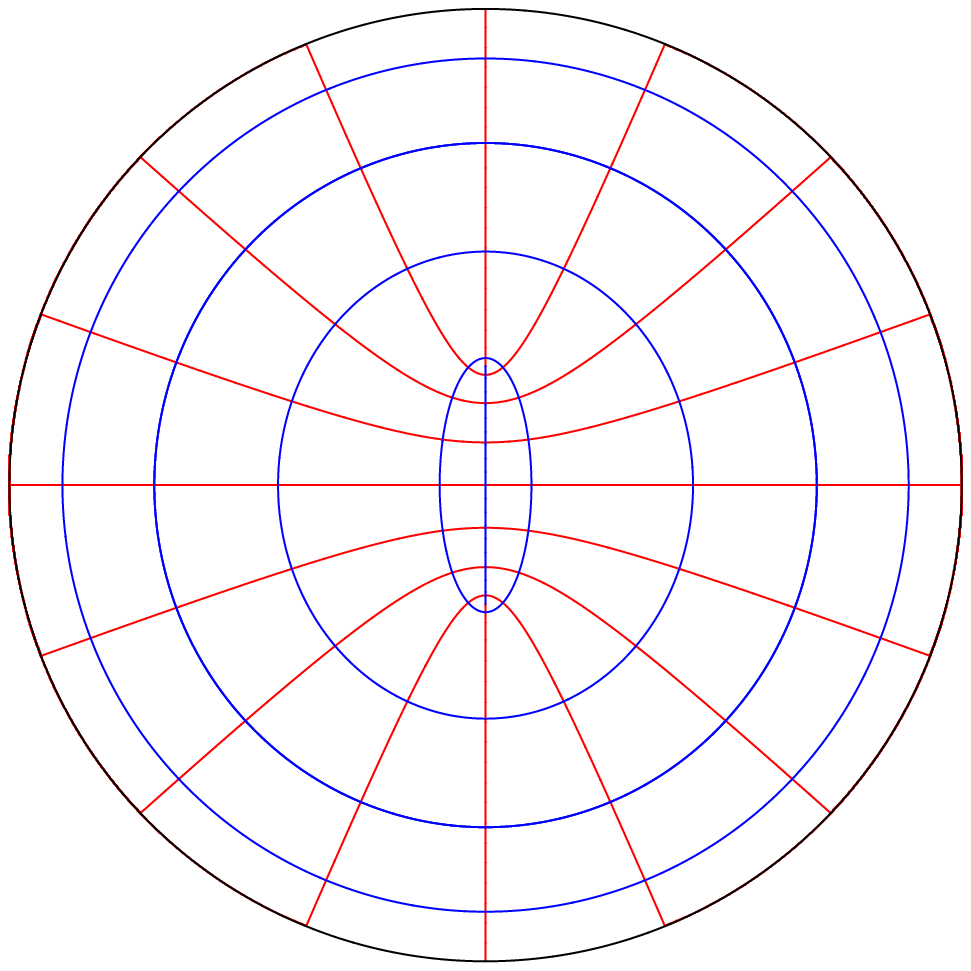}}
\caption{\label{sphero2} Coordinate curves $\xi=\mathrm{const.}$ and
  $\eta=\mathrm{const.}$}
\begin{picture}(1,1)(0,0)
\put(-10.5,233){\scriptsize $\xi=0$}
\put(-10.5,150){\scriptsize $\eta=0$}
\put(48,150){\scriptsize $\xi=\beta$}
\put(75,225){\scriptsize $\eta=\infty$}
\end{picture}
\end{figure}
As $\xi$ and $\eta$ vary, these describe a family of hyperbolae and
ellipses respectively. A plot of the coordinate curves in the $z-\rho$
plane is shown in Figure \ref{sphero2}, where for convenience we
allow $\rho$ to be negative.

Using equations (\ref{cylpolmet}) and (\ref{psdspher}), one can show
that the metric on $\Ht$ in pseudospheroidal coordinates is
\begin{equation}
\label{spheroidalmetric}
ds^2 = \frac{\xi^2+\eta^2}{(\alpha^2-\xi^2)(\beta^2-\xi^2)} d\xi^2 +
\frac{\xi^2+\eta^2}{(\alpha^2+\eta^2)(\beta^2+\eta^2)}d\eta^2 + \frac{\xi^2\eta^2}{\alpha^2\beta^2}d\phi^2
\end{equation}

\subsubsection{Pseudoparabolic Coordinates}
\label{ppara}

For one-centre problems in flat space with a distinguished spatial
direction, it is often useful to use parabolic coordinates. These may
be thought of as spheroidal coordinates where one of the two centres
has been allowed to recede to infinity, while the other is fixed at
the origin. Again following Vozmischeva \cite{vosm}, we shall
construct an analogous set of coordinates on $\Ht$ from the
pseudospheroidal coordinates given above.

We first apply a Lorentz transformation to $\mathbb{E}^{3,1}$ in order
to move the centre at $(\alpha, 0, 0, -\beta)$ to $(1,0,0,0)$. Without
loss of generality we may write $\alpha = \cosh \frac{\psi}{2}$,
$\beta = \sinh \frac{\psi}{2}$. The
point at $(W,Z,P,\phi)$ moves to $(W', Z', P', \phi')$, where the new
coordinates are related to the old by:
\begin{eqnarray}
\nonumber W &=& \cosh \frac{\psi}{2} W' - \sinh \frac{\psi}{2} Z', \\
\nonumber Z &=& \cosh \frac{\psi}{2} Z' - \sinh \frac{\psi}{2} W', \\
P&=& P', \qquad \quad \phi = \phi'. \label{lorentz}
\end{eqnarray}
We also rescale the variables $\xi$ and $\eta$ by
\begin{equation}
\xi = \alpha \xi' = \cosh \frac{\psi}{2} \xi', \qquad \eta = \beta
\eta' = \sinh\frac{\psi}{2} \eta' \label{rescale}
\end{equation}
In order to find the new coordinate surfaces, we substitute
(\ref{lorentz}) and (\ref{rescale}) into (\ref{cordsurf}) and take the
limit as $\psi \to \infty$, keeping terms of order $e^{-\psi}$. The
new coordinate surfaces are then given by:
\begin{eqnarray}
2Z'(W'-Z') &=& \frac{\xi'^2}{1-\xi'^2}(W'-Z')^2 -
\frac{1-\xi'^2}{\xi'^2}P'^2 = \mu^2(W'-Z')^2-\frac{P'^2}{\mu^2} \label{musurf}\\
2Z'(W'-Z') &=& -\frac{\eta'^2}{1+\eta'^2}(W'-Z')^2 +
\frac{1+\eta'^2}{\eta'^2}P'^2 = -\nu^2(W'-Z')^2+\frac{P'^2}{\nu^2} \label{nusurf}
\end{eqnarray}
Where we have defined new coordinates $\mu = \frac{\xi'^2}{1-\xi'^2}$
and $\nu = \frac{\eta'^2}{1+\eta'^2}$. We now have no further use for
the unprimed coordinates, so drop the primes. We can solve equations
(\ref{musurf}), (\ref{nusurf}), together with the pseudosphere
constraint $W^2-P^2-Z^2=1$ to express the points on the pseudosphere
in $\mathbb{E}^{3,1}$ in terms of $\mu$ and $\nu$. This gives
\begin{equation}
W = \frac{2+\mu^2-\nu^2}{2 \sqrt{(1-\nu^2)(1+\mu^2)}},\quad
P=\frac{\mu\nu}{\sqrt{(1-\nu^2)(1+\mu^2)}}, \quad Z =
\frac{\mu^2-\nu^2}{2 \sqrt{(1-\nu^2)(1+\mu^2)}}, \label{e31munu}
\end{equation}
\begin{figure}[!t]
\centering \framebox {\includegraphics[height=3in,
width=3in]{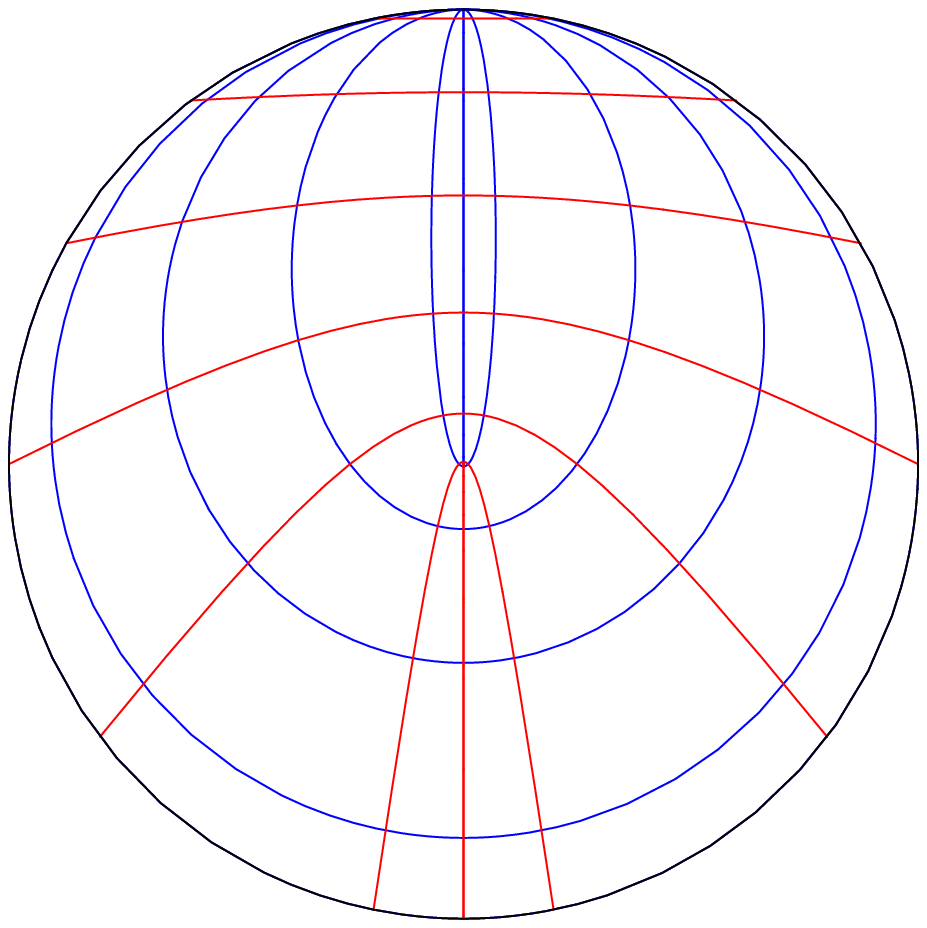}}
\caption{\label{para} Coordinate curves $\mu=\mathrm{const.}$ and
  $\nu=\mathrm{const.}$}
\begin{picture}(1,1)(0,0)
\put(-10.5,250){\scriptsize $\mu=\infty$}
\put(-10.5,190){\scriptsize $\nu=0$}
\put(-10.5,42){\scriptsize $\mu=0$}
\put(75,220){\scriptsize $\nu=1$}
\end{picture}
\end{figure}
where
\begin{equation}
\label{munurange}
0 \leq \mu < \infty, \qquad 0 \leq \nu < 1.
\end{equation}
This time the coordinates cover the whole of $\Ht$. The sphere at
infinity is given by $\nu =1$. We may once again use the gnomonic projection in order to visualise
the coordinate surfaces as surfaces within the Beltrami ball model. We
find that the Beltrami coordinates $z, \rho$ are given in terms of
$\mu, \nu$ by:
\begin{equation}
z = \frac{Z}{W}=\frac{\mu^2-\nu^2}{2+\mu^2-\nu^2}, \qquad \rho =
\frac{P}{W} = \frac{\mu\nu}{2+\mu^2-\nu^2}.
\end{equation}
Either from these equations, or from the projected versions of
equations (\ref{musurf}), (\ref{nusurf}), we find that the surfaces
$\mu = \mathrm{const.}$ are hyperboloids and those of $\nu =
\mathrm{const.}$ are ellipsoids. The plane $\phi=0$ is shown in Figure
\ref{para}.

Using equations (\ref{e31munu}) and (\ref{cylpolmet}), it is a matter
of straightforward calculation to find the metric on $\Ht$ in the $\mu, \nu,
\phi$ coordinates. We find that
\begin{equation}
\label{paraboloidalmetric}
ds^2 = \frac{\mu^2+\nu^2}{(1-\nu^2)(1+\mu^2)} \left [ \frac{d\mu^2}{1+\mu^2}+\frac{d\nu^2}{1-\nu^2}+\frac{d\phi^2}{\mu^{-2}+\nu^{-2}} \right ]
\end{equation}

\subsection{Horospheres \label{horo}}

\begin{figure}[!ht]
\[
\input{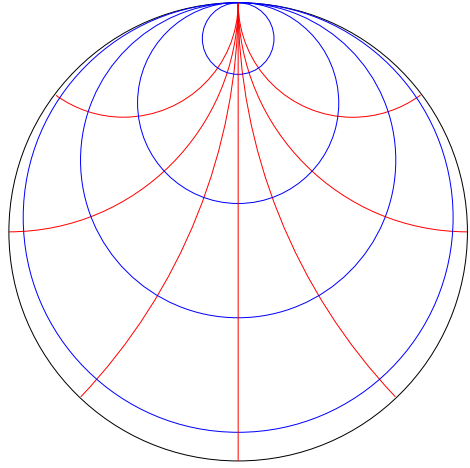}
\]
\caption{A plot showing horospheres and their normal geodesics in
  Poincar\'{e} coordinates}
\label{hor}
\end{figure}

When considering scattering processes in Euclidean space, the incoming
wave is usually considered to be a plane wave, i.e. a wave whose phase
is constant along any plane parallel to some vector. The important
property of such planes is that they are everywhere normal to a set of
parallel geodesics. Whether we are considering classical action waves
or wavefunctions, the interpretation is that particles are moving
parallel to some fixed direction. For hyperbolic space, we would like
a similar set of waves to describe incoming particles initially
parallel to some direction. In order to do this, we consider a pencil of
parallel geodesics in $\Ht$, which meet at a point on $\partial
\Ht$. The horospheres are a set of surfaces everywhere normal to these
geodesics. It is most convenient to visualise these surfaces in the
Poincar\'{e} ball model, since this is conformally flat so angles are
as one would expect in Euclidean space. A set of parallel geodesics in this model are
a set of circular arcs which meet the sphere $r=1$ orthogonally at
some given point, say $(0,0,1)$. A surface which is everywhere normal
to this pencil of geodesics is a sphere which is tangent to the sphere
$r=1$ at the point $(0,0,1)$. A picture of the $y=0$ plane is shown in
figure \ref{hor}.

The horospheres may be thought of as a limit of a set of spheres in
$\Ht$ whose centre goes to infinity whilst the radius also tends to
infinity. The horospheres carry a natural Euclidean geometry and the
restriction of the hyperbolic metric to each horosphere is flat, a
result known as Wachter's theorem
\cite{Fenchel, Coxeter}. Clearly there is nothing special about our choice of boundary point,
so there is a set of horospheres associated with every point on
$S^2_\infty = \partial \Ht$.

A brief calculation shows that the horospheres associated with a
pencil of parallel geodesics originating from $(0,0,1)$ are given in
pseudoparabolic coordinates by the equation
\begin{equation}
(1+\mu^2)(1-\nu^2) = \mathrm{const.}
\end{equation}

\end{document}